\documentclass[pra,english,superscriptaddress,nofootinbib,twocolumn]{revtex4-1}
\usepackage[T1]{fontenc}
\usepackage[latin9]{inputenc}
\usepackage[tbtags]{amsmath}
\usepackage{hyperref}
\usepackage{tikz}
\usetikzlibrary{matrix}
\usepackage[toc,page]{appendix}
\usepackage{braket}
\usepackage{cancel}
\usepackage{tabu}
\usepackage{array}
\usepackage{graphicx,color}
\usepackage{amsfonts}
\usepackage{multirow}

\usepackage{tabularx}
\usepackage[thicklines]{easytable}
\DeclareMathOperator{\Tr}{Tr}

\makeatletter
\newcommand\numberthis{\addtocounter{equation}{1}\tag{\theequation}}
\usepackage{babel}

\makeatother
\usepackage{babel}

\begin{document}

\title{Prethermalization of quantum systems interacting with non-equilibrium environments}

\author{Andreu Angl\'es-Castillo}

\affiliation{Departament de F\'isica Te\`orica \& IFIC, Universitat de Val\`encia-CSIC,
46100 Burjassot (Val\`encia) Spain}

\author{Mari Carmen Ba\~nuls}

\affiliation{Max-Planck-Institut für Quantenoptik, Hans-Kopfermann-Str. 1, 85748
Garching, Germany}

\affiliation{Munich Center for Quantum Science and Technology (MCQST), Schellingstr. 4,
D-80799 München}

\author{Armando P\'erez}

\affiliation{Departament de F\'isica Te\`orica \& IFIC, Universitat de Val\`encia-CSIC,
46100 Burjassot (Val\`encia) Spain}

\author{In\'es De Vega}

\affiliation{Department of Physics and Arnold Sommerfeld Center for Theoretical
Physics, Ludwig-Maximilians-University Munich, Germany}

\begin{abstract}

The usual paradigm of open quantum systems falls short when the environment is actually coupled to additional fields or components that drive it out of equilibrium. Here we explore the simplest such scenario, by considering a two level system coupled to a first thermal reservoir that in turn couples to a second thermal bath at a different temperature. We derive a master equation description for the system and show that, in this situation, the dynamics can be especially rich. In particular, we observe prethermalization, a transitory phenomenon in which the system initially approaches thermal equilibrium with respect to the first reservoir, but after a longer time converges to the thermal state dictated by the temperature of the second environment. Using analytical arguments and numerical simulations, we analyze the occurrence of this phenomenon, and how it depends on temperatures and coupling strengths. The phenomenology gets even richer if the system is placed between two such non-equilibrium environments. In this case, the energy current through the system may exhibit transient features and even switch direction, before the system eventually reaches a non-equilibrium steady state.

\end{abstract}

\maketitle

\section{Introduction}
The standard theory of open quantum systems (OQS) typically considers that the system is coupled to a single reservoir in equilibrium to analyse properties such as decoherence, dissipation and non-Markovianity \cite{rivas2014,breuer2015,devega2017,li2018}. A  richer situation emerges in the frame of quantum thermodynamics and thermal machines, in which the system is coupled to two or more reservoirs, each of them 
 equilibrated at a different temperature and/or chemical potential \cite{eitan1992,kosloff2013,schaller2014,goold2016}. Once the coupling is activated, the open system evolves towards a non-equilibrium steady state that may contain persistent heat, particle or spin currents. 
 An even more involved  scenario occurs when the system is coupled to one or more reservoirs that are each of them out of equilibrium and therefore evolve in time.  
 As a consequence, the action of the environment into the system dynamics is no longer encoded in a correlation function that is time-translational invariant, such that $\alpha(t,\tau)=\alpha(t-\tau)$, but rather on a correlation function that depends on both the current time of evolution $t$ and the past times $\tau$.
 
The motivation to analyze complex environments beyond the standard OQS paradigm of single and multiple equilibrium reservoirs is strong. From an application perspective, out of equilibrium environments that present a temperature gradient can be encountered in electron transfer processes in quantum chemistry and biology \cite{bio1}, in cellular media \cite{bio2} and even in the thermosynthesis processes that use the solar energy to create chemical compounds \cite{bio3}, to name just a few examples. These types of environments may indeed be driven by an external source, corresponding to other molecular or biological structures or even to the electromagnetic field. Non-equilibrium environments are also present in quantum technological devices, 
where the quantum system of interest may be directly coupled to an environment that is itself coupled to a second reservoir, thermalized at a different temperature. Such temperature gradient of the different components and subsystems surrounding the quantum system of interest is particularly present in quantum computers \cite{thermal-grad-qc,thermal-grad-trapped}. Superconducting qubits, for instance, are cooled down to cryostatic temperatures, while their surrounding components, including amplifiers and processing units, as well as the cables and waveguides that connect them to each other and to the qubit, are at higher temperatures the further they are from the circuit.

Describing these situations is of fundamental and timely interest, but it also represents a significant challenge, as the effects of indirect reservoirs on the OQS dynamics can not be captured with a simple Markovian approximation. To this aim, one possibility is to compute the full dynamics, including the system and the environments, and then trace out the environmental degrees of freedom to obtain the OQS reduced dynamics. However, the dimension of the full Hilbert space grows exponentially fast with the number of degrees of freedom, and further, the relevant states may be largely entangled as well, which makes inefficient a direct use of state of the art numerical methods like Monte Carlo \cite{metropolis,troyer2009,Bauer2011alps} and matrix product states \cite{Cirac2009rg,Verstraete2008,Schollwoeck2011,Orus2014a,Silvi2017tns}.

While several approaches can be found in the literature to describe the full dynamics of the system coupled to a single bath \cite{10.1063/1.3159671,PhysRevLett.105.050404,10.1063/1.3490188,PhysRevA.92.052116}, much fewer works can be found that treat the presence of a second environment driving the first one out of equilibrium.
In this context, Ref. \cite{StochasticEnvironment} considers an effective (surrogate) Hamiltonian to describe the system and its direct coupling to a primary environment (represented by a finite number of modes), while a second and larger environment is introduced and coupled to the first. This second environment is treated stochastically.
Here we propose an alternative approach,
which extends the standard tools of the OQS theory, namely the weak coupling approximation and the master equation approach, to consistently tackle the problem in at least a limit of interest.

To be specific, we consider a two-level quantum system coupled to a first reservoir (RI) that is in turn coupled to a second reservoir (RII). Initially, each reservoir is in a thermal state at a different temperature, respectively $T_\mathrm{I}$ and $T_\mathrm{II}$. 
We additionally consider that RII induces a Markovian evolution on the modes of RI so that they thermalize efficiently.
 Therefore, even if RI is initially in thermal equilibrium, the coupling to a second reservoir at a different temperature will drive it away from it, and enforce its evolution towards a new equilibrium state with respect to RII. Thus, the dissipation of the open system will display very rich features reflecting the interplay between two different timescales: 
thermalization of the system at a temperature $T_\mathrm{I}$, 
 and the thermalization to its final equilibrium state with $T_{\mathrm{II}}$. If the conditions of the environment are suitable, and these two timescales are temporally separated, prethermalization \cite{PrethermalizationOriginal} of the OQS is observed, which is a stage in which the system remains thermalized at $T_\mathrm{I}$. 

The plan of the paper is the following: We present the details of our model in Sec. \ref{II}, while in Sec. \ref{III} we discuss the master equation that is used to describe the reduced dynamics of the open system. This master equation depends on a set of correlation functions that encode the effects of both reservoirs in the open system, and which are discussed in Sec. \ref{IV}. Sections \ref{V} and \ref{VI} describe the effects of prethermalization when considering a single and two out of equilibrium reservoirs, respectively. Finally, we draw some conclusions in Sec. \ref{VII}.

\section{Model with two interacting environments}
\label{II}

\begin{figure}
\centering
\begin{tikzpicture}
\tikzstyle{S}=[draw,circle,fill=blue!50!white,text=black,minimum width=35pt]
\tikzstyle{RI}=[draw,circle,fill=red!50!white,text=black,minimum width=20pt]
\tikzstyle{RII}=[draw,circle,fill=green!30!white,text=black,minimum width=5pt]
  \foreach \place/\name in {{(0,0)/S}}
    \node[S] (\name) at \place {\LARGE \name};
    
  \foreach \place/\name/\cosa in {{(0,2)/a/$a_1$},  {({2*cos(30)},{2*sin(30)})/c/$a_2$}, {({2*cos(60)},-{2*sin(60)})/d/$a_\lambda$}}
    \node[RI] (\name) at \place {\cosa};
    
  \foreach \source/\dest in {S/c, S/d, S/a}
    \path (\source) edge (\dest);
    
  \foreach \place/\name/\cosa in {{(0,3)/e/$b_{11}$},{({0+cos(30)},{2+sin(30)})/g/$b_{12}$},{({0+cos(60)},{2-sin(60)})/h/$b_{1k}$}}
    \node[RII] (\name) at \place {\tiny \cosa};
    
  \foreach \source/\dest in {a/e, a/g, a/h}
    \path (\source) edge (\dest);

  \foreach \place/\name/\cosa in {{({2*cos(30)},{2*sin(30)+1})/i/$b_{21}$},{({2*cos(30)+cos(30)},{2*sin(30)+sin(30)})/j/$b_{22}$},{({2*cos(30)+cos(60)},{2*sin(30)-sin(60)})/k/$b_{2k}$}}
    \node[RII] (\name) at \place {\tiny \cosa};
    
  \foreach \source/\dest in {c/i, c/j, c/k}
    \path (\source) edge (\dest);
    
  \foreach \place/\name/\cosa in {{({2*cos(60)},{2*sin(-60)+1})/l/$b_{\lambda 1}$},{({2*cos(60)+cos(30)},{2*sin(-60)+sin(30)})/m/$b_{\lambda 2}$},{({2*cos(60)+cos(60)},{2*sin(-60)-sin(60)})/n/$b_{\lambda k}$}}
    \node[RII] (\name) at \place {\tiny \cosa};
    
  \foreach \source/\dest in {d/l, d/m, d/n}
    \path (\source) edge (\dest);

  \draw[dotted,thick] ({2*cos(105)},{2*sin(105)}) arc (105:(360-75):2);
  \draw[dotted] ({0+cos(0)},{2+sin(0)}) arc (0:-30:1);
  \draw[dotted] ({0+cos(120)},{2+sin(120)}) arc (120:270:1);
  \draw[dotted] ({2*cos(30)+cos(0)},{2*sin(30)+sin(0)}) arc (0:-30:1);
  \draw[dotted] ({2*cos(30)+cos(220)},{2*sin(30)+sin(220)}) arc (220:270:1);
  \draw[dotted] ({2*cos(60)+cos(0)},{2*sin(-60)+sin(0)}) arc (0:-30:1);
  \draw[dotted] ({2*cos(60)+cos(120)},{2*sin(-60)+sin(120)}) arc (120:270:1);  
\end{tikzpicture}
	\caption{Schematic picture of the model. The OQS is coupled in a star configuration to a set of harmonic oscillators $a_\lambda$, which are coupled to their own individual baths of harmonic oscillators $b_{\lambda, k}$.}
	\label{fig:ModelB}
\end{figure}
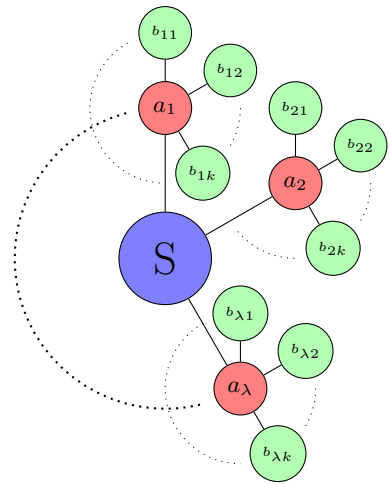

As is standard in the theory of OQS \cite{Breuer,OpenQuantumSystemsIntro,devega2017} we consider that the total evolution of system plus the environment is unitary and described by the Hamiltonian, 
\begin{equation}
	H=H_\textrm{S}+H_\textrm{E}+H_{\textrm{int}}~,
\end{equation}
where $H_\textrm{S}$ and $H_\textrm{E}$ are the free Hamiltonians of the system and environment, respectively, and $H_{\textrm{int}}$ is the interaction Hamiltonian between system and environment.
We model the system as a two level system with the free Hamiltonian
\begin{equation}\label{eqn:HamiltonianSystem} 
	H_\textrm{S}=\frac{1}{2}\omega_0 \sigma_z~,
\end{equation}
where $\omega_0$ is the energy\footnote{Throughout this article we consider natural units in which the reduced Plank constant and the Boltzmann constant $\hbar=k_B=1$.} gap between levels. 
We model the environment to which the system is coupled as an 
set of open harmonic oscillators that is a first reservoir (RI) of harmonic oscillators where each mode in RI is coupled to a independent reservoir, included in the second reservoir (RII). 
The Hamiltonian describing this environment is
\begin{equation}
	H_\textrm{E}= H_{\textrm{RI}} + H_{\textrm{RII}} + H_{\textrm{int},2}~,
\end{equation}
where
\begin{equation}
	H_{\textrm{RI}} = \sum_\lambda \omega_\lambda a_\lambda^\dagger a_\lambda \quad \text{and} \quad H_{\textrm{RII}}=\sum_{\lambda,k} \omega_{\lambda,k} b_{\lambda,k}^\dagger b_{\lambda,k}~,
\end{equation}
are the free Hamiltonians of RI and RII, respectively, with operators that obey the commutation relations
\begin{equation}\label{eqn:CommutationRelations}
 [a_\lambda,a_ {\lambda'}^\dagger]=\delta_{\lambda,\lambda'}~, \quad  \text{and} \quad      [b_{\lambda k},b_{\lambda' k'}^{\dagger}]=\delta_{\lambda,\lambda'}\delta_{k,k'} ~,
\end{equation} and whose interaction
\begin{equation}\label{eqn:HamEnvirInter}
	H_{\textrm{int},2} = \sum_\lambda \left( a_\lambda^\dagger \otimes \sum_k \tilde{g}_{\lambda,k} b_{\lambda,k} + a_\lambda  \otimes \sum_k\tilde{g}_{\lambda,k}^*b_{\lambda,k}^\dagger\right) ~,
\end{equation}
conserves the boson number. The coupling strength between the $\lambda$-th oscillator in RI and the $k$-th oscillator in RII is $\tilde{g}_{\lambda,k}$. 
The system is in a \textit{star configuration}, i.e. the OQS is coupled to all the $\lambda$ bosonic operators of RI, and in turn each of these is coupled to a reservoir of harmonic oscillators that is a part of RII, as depicted in Fig.~\ref{fig:ModelB}. 
Only the first reservoir couples directly to the OQS, with the interaction Hamiltonian
\begin{equation}\label{eqn:HamInter}
	H_{\textrm{int}}= \sigma_- \otimes \sum_\lambda g_\lambda^* a_\lambda^\dagger + \sigma_+ \otimes \sum_\lambda g_\lambda a_\lambda ~,
\end{equation}
which only considers interactions that conserve the particle number. We take as initial state a tensor product,
\begin{equation}\label{eqn:InitialState}
	\rho(0)=\rho_\textrm{S}(0) \otimes \rho_\textrm{E}(0)=\rho_\textrm{S}(0) \otimes \rho_{\textrm{I}}^{\textrm{th}}(\beta_\textrm{I}) \otimes \rho_{\textrm{II}}^{\textrm{th}}(\beta_\textrm{II})~.
\end{equation}
The initial state of the system can be arbitrary, while the initial states of the reservoirs are assumed to be thermal, possibly at different temperatures, 
\begin{equation}\label{eqn:InitialThermal}
	\rho_{i}^{\textrm{th}}(\beta_i) = \frac{e^{-\beta_i H_{\textrm{R}i}}}{Z_i(\beta_i)},
\end{equation}
where  $Z(\beta_i) = \text{Tr} \{e^{-\beta_i H_{\textrm{R}i} }\}$ is  the partition function and $\beta_i=1/T_i$ is the inverse temperature of each reservoir,  $i=\{\textrm{I},\textrm{II}\}$.

Each reservoir may have a different spectral function depending on their microscopic properties and the problem considered. In our analysis, we consider the Caldeira-Leggett phenomenological model of spectral functions \cite{RevModPhys.59.1}, which reads
\begin{equation}\label{eqn:SpectralDens}
	J_i(\omega)=g_i \omega_{ci}^{1-s_i}\omega^{s_i} e^{-\omega/\omega_{ci}}~,
\end{equation}
where $g_i$ is the strength of the coupling, $s_i$ is a factor that takes different values depending on the particular environment that needs to be modelled, and $\omega_{ci}$ determines a smooth frequency cut off for the modes of the reservoir.

\section{Study of the system evolution}
\label{III}

To obtain a closed equation for the dynamics of the open quantum system, we consider that it is weakly coupled to its environment, which makes it evolve slowly. Thus we can derive a second order weak coupling master equation (ME) for the reduced density matrix of the open quantum system. The derivation of the ME is standard and can be found in numerous works \citep{PhysRevA.95.052108,Breuer},
\begin{equation}\label{eqn:ME}
\begin{split}\frac{d}{dt}\rho_{\textrm{S}}(t)= & -i\left[H_{\textrm{S}},\rho_{\textrm{S}}(t)\right]\\
 & +\Bigg  ( \int_{0}^{t}d\tau\alpha^{+}(t,\tau)\left[V_{\tau-t}\sigma_+\rho_{\textrm{S}}(t),\sigma_-\right]\\
 & +\int_{0}^{t}d\tau\alpha^{-}(t,\tau)\left[V_{\tau-t}\sigma_-\rho_{\textrm{S}}(t),\sigma_+\right] + h.c. \Bigg)~,
\end{split}
\end{equation}
where $V_{t}\mathcal{O}=e^{iH_{\textrm{S}}t}\mathcal{O}e^{-iH_{\textrm{S}}t}$ is the free evolution of
the operator $\mathcal{O}=\{\sigma_+,\sigma_-\}$, and the correlation functions are defined by 

\begin{equation}
\begin{split}\alpha^{+}(t,\tau) & =\text{Tr}\{B(t)^{\dagger}B(\tau){\rho}_{\textrm{E}}(0)\}~,\\
\alpha^{-}(t,\tau) & =\text{Tr}\{B(t)B^{\dagger}(\tau){\rho}_{\textrm{E}}(0)\}~,
\end{split}
\label{eqn:CorrelDef}
\end{equation}
with $B(t)=e^{iH_{\textrm{E}}t}Be^{-iH_{\textrm{E}}t}$ the free evolution of the environment
operator $B=\sum_\lambda g_\lambda a_\lambda$.
Notice that this equation is second order in the interaction
operator $B$, and that no first order term is present, since it is proportional to $\text{Tr}_E\{H_{\textrm{int}}(t) \rho_\textrm{E}(0)\}$, which is null for the initial state defined in Eqs.~(\ref{eqn:InitialState},\ref{eqn:InitialThermal}). This equation is a time-local ME, since
its evolution can be recast in the form 
\begin{equation}
\dot{\rho}_{\textrm{S}}(t)=\Lambda_{t}[\rho_{\textrm{S}}(t)]~,\label{eqn:timelocalME}
\end{equation}
where $\Lambda_{t}$ is a linear map, 
such that $\Lambda_t [\rho(t)]$ is Hermitian and traceless for any $\rho$. To fully describe
the OQS through the differential equation (\ref{eqn:ME}), the correlation
functions (\ref{eqn:CorrelDef}) have to be computed for the initial
states $\rho_{\textrm{E}}(0)$ defined in Eqs.~(\ref{eqn:InitialState},\ref{eqn:InitialThermal}).
The following section is devoted to this derivation, but first we rewrite the ME in Eq.~(\ref{eqn:ME}) under its canonical form.

\subsection{Canonical Form of the ME}

Any time-local ME equation of the form (\ref{eqn:timelocalME}) can
be recast into a canonical ME \cite{PhysRevA.89.042120}, of the form
\begin{equation}
\begin{split} & \frac{d}{dt}\rho_{\textrm{S}}(t)=-i[H(t),\rho_{\textrm{S}}(t)]+\\
 & \sum_{k=1}^{d^{2}-1}\gamma_{k}(t)\Big(L_{k}(t)\rho_{\textrm{S}}(t)L_{k}^{\dagger}(t)-\frac{1}{2}\{L_{k}^{\dagger}(t)L_{k}(t),\rho_{\textrm{S}}(t)\}\Big)~,
\end{split}
\end{equation}
where $\gamma_{k}(t)$ are the canonical decay rates corresponding
to the canonical decoherence channels $L_{k}(t)$, with $k=1$, \dots, $d^{2}-1$,
and $d$ the dimension of the Hilbert space of the OQS. $H(t)$ is, in general, not identical
to the free Hamiltonian of the system, since the interaction with
the environment modifies it. The most common effect
is a shift of the natural frequency of the OQS, the so-called Lamb
shift.
The equation is often written in a more compact form as
\begin{equation}\label{eqn:CanonicalME}
	\frac{d}{dt}\rho_{\textrm{S}}(t)=-i[H(t),\rho_{\textrm{S}}(t)]+\mathcal{D}(t,\rho_\textrm{S}(t))~.
\end{equation} 
where the first term represents the unitary evolution of the OQS. The second term in \eqref{eqn:CanonicalME} encompasses the dissipative part of the evolution. 

Recasting the time-local ME in this form allows us to easily evaluate whether, despite being an approximated equation, it still  preserves complete positivity of the evolution. In detail, if the decay rates $\gamma_{k}(t)$ are non-negative we can ensure that this is the case and that the dynamical map of the OQS is Markovian \cite{PhysRevA.89.042120}. The canonical decay rates, and the Lamb shift for our model, are discussed in the next section and in Appendix \ref{append:ME}.

\section{Out-of-equilibrium correlation functions and decay rates}
\label{IV}

Obtaining the correlation functions (\ref{eqn:CorrelDef}) requires to compute the time evolution of $a_\lambda(t)$ in the operator $B(t)=\sum_{\lambda}g_{\lambda}a_{\lambda}(t)$. We can simplify this calculation by assuming a large separation of timescales between the second and the first reservoir. Specifically, we consider that the modes of the first reservoir, $a_\lambda(t)$ slowly evolve towards an equilibrium state with respect to the second reservoir, and that this evolution is well described with the Markov approximation. This is discussed in Appendix \ref{append:OperatorA}, while the computation of the correlation functions is treated in Appendix \ref{append:correlME}.
Thus, the correlation functions are given by
\begin{equation}\label{eqn:CorrelationFunctions}
\begin{split}
	\alpha^{+}(t,t')=&\frac{1}{\pi}  \int d\omega J_{\textrm{I}}(\omega)n_{\textrm{I}}(\omega)e^{i\omega t'}e^{-		\frac{J_{\textrm{II}}(\omega)}{2}(2t-t')}\\
	+\frac{1}{\pi^{2}}\iint d&\omega d\omega'  J_{\textrm{I}}(\omega)n_{\textrm{II}}(\omega') K(\omega,\omega')C(\omega,\omega',t,t')~,\\
	\alpha^{-}(t,t')=&\frac{1}{\pi}  \int d\omega J_{\textrm{I}}(\omega)(n_{\textrm{I}}(\omega)+1)e^{-i\omega t'}e^{-\frac{J_{\textrm{II}}(\omega)}{2}(2t-t')}\\
	+\frac{1}{\pi^{2}}\iint d&\omega d\omega'  J_{\textrm{I}}(\omega)(n_{\textrm{II}}(\omega')+1)K(\omega,\omega')C^*(\omega,\omega',t,t')~,
\end{split}
\end{equation}
where $J_{i}(\omega)$, with $i=\{\textrm{I},\textrm{II}\}$, are the spectral
functions of each reservoir, which have the general form (\ref{eqn:SpectralDens}),
and $n_i (\omega)=[\exp(\beta_i\omega) - 1]^{-1}$
is the average thermal number of quanta in mode $\omega$ at inverse
temperature $\beta_{i}$. We have defined the function 
\begin{equation}\label{eqn:Kernel}
	K(\omega,\omega')=\frac{J_{\textrm{II}}(\omega')}{\left(\frac{J_{\textrm{II}}(\omega)}{2}\right)^{2}+(\omega-\omega')^{2}}~,
\end{equation} 
which is proportional to a Lorentzian kernel of width  $J_{\textrm{II}}(\omega)/2$,
and the function
\begin{equation}
\begin{split}
	C(\omega,\omega',t,t')=&\left[e^{-i\omega't}-e^{\left(-i\omega-\frac{J_{\textrm{II}}(\omega)}{2}\right)t}\right] \times \\ 
	&\left[e^{i\omega'(t-t')}-e^{\left(i\omega-\frac{J_{\textrm{II}}(\omega)}{2}\right)(t-t')}\right]
\end{split}~.
\end{equation}
Notice that, even though we can consider that the open system is weakly coupled to RI, and thus its master equation is obtained within a second order perturbation theory, a Markov approximation can not be taken in a straightforward way. The reason is that the correlation functions  (\ref{eqn:CorrelationFunctions}) are no longer dependent on the time difference $t-\tau$, but on both times $t$ and $\tau$ such that one can not simply extend the integration limits in Eq. (\ref{eqn:ME}) by assuming that the integral kernel decays much faster than the system evolution time-scale, as it is done in the Markov approximation.

We observe that the second term of the correlation functions contains the resonant term $K(\omega,\omega')$ (see Eq.~(\ref{eqn:Kernel})) with a width proportional to the coupling strength between environments, and centered at $\omega=\omega'$. Approximating this term by a delta function is consistent with the weak coupling approximation already considered between RI and RII.
Using this approach, we obtain an analytical approximation for the canonical decay rates $\gamma_{\pm}(t)$,
which correspond to the decoherence channels $L_{\pm}=\sigma^{\pm}$ (see appendix~ \ref{append:ME}), and which can be split into two contributions, $\gamma_{\pm}(t)=\gamma_{\pm}^{\textrm{ST}}(t)+\gamma_{\pm}^{\textrm{LT}}(t)$,
 where the terms are labelled in reference to their short time (ST) or long time (LT) dominance.
 The ST terms are
\begin{equation}\label{eqn:ApproxDecayRates1}
\begin{split}
	\gamma_+^{\textrm{ST}}(t)&=J_\textrm{I}(\omega_0)n_\textrm{I}(\omega_0)e^{- J_{\textrm{II}}(\omega_0) t}\\
	\gamma_-^{\textrm{ST}}(t)&= J_\textrm{I}(\omega_0)(n_\textrm{I}(\omega_0)+1)e^{- J_{\textrm{II}}(\omega_0) t}~,
\end{split}
\end{equation}
and the LT terms read
\begin{equation}\label{eqn:ApproxDecayRates2}
\begin{split}
	\gamma_+^{\textrm{LT}}(t)&= J_\textrm{I}(\omega_0) n_{\textrm{II}}(\omega_0)(1-e^{- J_{\textrm{II}} (\omega_0) t})~, \\
	\gamma_-^{\textrm{LT}}(t)&= J_\textrm{I}(\omega_0) (n_{\textrm{II}}(\omega_0)+1) (1-e^{- J_{\textrm{II}} (\omega_0) t})~.
\end{split}
\end{equation}
The validity of approximating Eq.~(\ref{eqn:Kernel}) by a delta function is discussed in Appendix \ref{append:ApproxDecay}. These decay rates present a very suggestive form: at short times, the LT terms of each decay rate is negligible, while at later times it dominates (see Appendix \ref{append:ApproxDecay} for a visual reference). 
The strength of the decay rates is governed by the spectral function of the first environment, while the second environment spectral function is responsible for the timescales at which each term dominates. 

With this approximate expression for the decay rates it is possible to prove analytically that indeed the OQS evolves, at long times, to a thermal state at the inverse temperature of the second reservoir $\beta_{\mathrm{II}}$ (see appendix~\ref{append:Asymptotic}). Furthermore, since they are non-negative at all times, we can ensure that the ME preserves complete positivity.

\begin{figure*}
\centering
	\includegraphics[trim=7 6.7 0 0,clip,width=\linewidth]{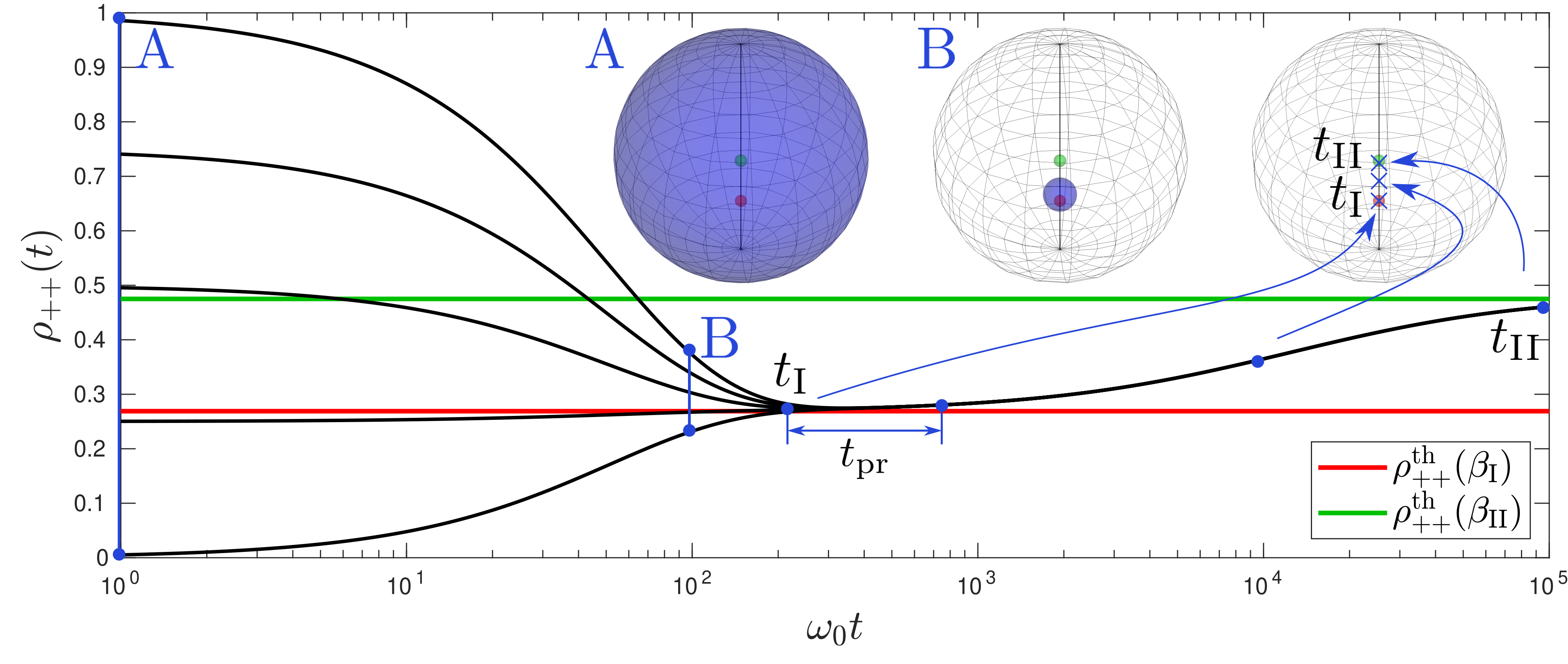}
	\caption{Evolution of the population of the $\ket{+}$ state for different initial pure states, and snapshots of the evolution of the ball of accessible states. Initially, all initial conditions tend to the upper population of the thermal state with $\beta_\textrm{I}$ at time $t_\textrm{I}$, from which point the evolution is identical. This effect translates into the reduction of the ball of accessible states to a point around the thermal state at $\beta_\textrm{I}$, as shown in Eq. (\ref{pointbloch}), represented by a red dot. The OQS stays close to this thermal state for some time: The prethermalization time $t_{\textrm{pr}}$. Afterwards, the OQS starts the evolution towards the thermal state at $\beta_{\textrm{II}}$. This corresponds to the displacement of a point (marked as a cross) from the red dot to the green one (representing the thermal state at $\beta_{\textrm{II}}$). The arrows connect the points of the upper population with the snapshots of the evolution of the ball of accessible states. The environments parameters are $g_\textrm{I}=10^{-2}$, $g_{\textrm{II}}=10^{-5}$, $s_\textrm{I}=s_{\textrm{II}}=1$, $\omega_{c\textrm{I}}=\omega_{c\textrm{II}}=10$, $\beta_\textrm{I}=1$ and $\beta_{\textrm{II}}=0.1$ and the system frequency is $\omega_0=1$.}
	\label{fig:Pretherm}
\end{figure*}

\section{Prethermalization}
\label{V}

The decay rates obtained in the previous section already suggest that the evolution of
the OQS may exhibit two different timescales. However, as it will soon become apparent, how well defined these two scales are will be strongly determined not only by the value of $g_{II}$, but also by other factors, such as the temperature of each reservoir. Depending on these factors, there may be a transitory state in which the OQS remains close to a thermal state corresponding to the initial temperature of RI, but after some longer time it
finally relaxes to a thermal state with the temperature of RII. 

This transient effect is an instance of \textit{prethermalization}, a phenomenon in which the system, after a short time, seems to relax to a state different from the true thermal equilibrium, which is eventually reached after a much longer timescale \cite{Rigol-PhysRevX.9.021027,PrethermalizationOriginal, ReviewPrethermalization, GGE2,Mallayya2019}. 
The most studied scenario of prethermalization concerns weakly non-integrable systems, in which an eigenstate of an integrable model is evolved under a quenched Hamiltonian that weakly breaks integrability.
The short time dynamics is still determined by almost conserved quantities, and the system arrives to a prethermalized state, but at long times the breaking of the integrability dominates and the system finally thermalizes \cite{Integrability}.
The phenomenon has also been studied in the context of OQS in \cite{PrethermOQS1-PhysRevB.97.165138,PrethermOQS2-PhysRevB.97.024302} and observed experimentally in ultra-cold bosonic atoms \cite{Ultracold1,Ultracold2,Ultracold3}. 

In our setup, a small coupling to the second reservoir ($g_{\mathrm{II}}\neq 0$) can play a similar role to the integrability breaking, as it perturbs the thermal equilibrium of the enviroment RI (which would otherwise remain stable).
In this way, the initial temperature of RI may determine the short time evolution and the arrival to a prethermal state, while the  final equilibrium is determined by RII. 
We will thus consider that prethermalization has occurred when the system reaches a state, independent of its initial conditions, close to the thermal equilibrium at $\beta_{\mathrm{I}}$, and this state is mantained
for a finite time, before the evolution definitely drives the system  to the equilibrium with RII.

In order to verify the occurrence of the effect, we analyse the evolution of all possible initial states. 
We conveniently express the density matrix 
in terms of the polarization vector, $\rho(t)=(I+\vec{p}(t)\cdot \vec{\sigma})/2$ and integrate the time evolution equations (see Appendix \ref{append:ME}). The formal solution for the polarization vector is
\begin{equation}
	\vec{p}(t) = r(t) R(t) \vec{p}(0) + \vec{d}(t)~,
\end{equation}
where
\begin{equation}\label{eqn:RotationMatrix}
R(t)=
\begin{pmatrix}
		\cos (\tilde{\Omega}(t)) & -\sin (\tilde{\Omega}(t)) & 0 \\
		\sin (\tilde{\Omega}(t)) & \cos (\tilde{\Omega}(t)) & 0\\
		0 & 0 & 1
	\end{pmatrix}~,
\end{equation}
is a rotation matrix, that performs a rotation about the $z$ axis with angular frequency $\tilde{\Omega}(t)=\int_0^t dt' \Omega(t')$, where $\Omega(t)$ is 
the shifted frequency of the OQS due to the action of the environment (see Appendix \ref{append:ME}), $r(t)=e^{-\tilde{\Gamma}(t)}$ is a scaling factor that affects equally all components, with $\tilde{\Gamma}(t)=\int_0^t dt' (\gamma_+(t')+\gamma_-(t'))$, where $\gamma_+(t)$ and $\gamma_-(t)$ are the canonical decay rates, and $\vec{d}(t)=(0,0,c(t))$ is a displacement vector in the $z$ direction, with 
\begin{equation}
	c(t)=e^{-\tilde{\Gamma}(t)}\int_0^t dt' e^{\tilde{\Gamma}(t')}(\gamma_+(t')-\gamma_-(t'))~.
\end{equation}

From this result, it is apparent that the effect of the dynamical map on any state  is to rotate the polarization vector around the $z$ axis,
rescale it by $r(t)$ and add a displacement $c(t)$ along the vertical direction. These transformations are independent of the initial state, hence the space of accessible states, initially described by the volume limited by the Bloch sphere, is isotropically contracted and shifted, and can be characterized
by its time dependent radius and center.

We would like to emphasize that the Lamb shift does not play any role in the evolution of the diagonal elements of the reduced density matrix, which ultimately means that it does not affect either the long time thermalization or the prethermalization dynamics. It is encoded in the angular frequency of Eq.~(\ref{eqn:RotationMatrix}) and thus it has the effect of rotating the ball of accessible states with an angular velocity different from $\omega_0$, but does not affect the rescaling and displacement of the whole space.

This representation allows us to understand how fast the memory of the initial state is lost, and in which state the OQS is. 
For the approximate decay rates Eqs.~(\ref{eqn:ApproxDecayRates1},\ref{eqn:ApproxDecayRates2}) we obtain the following expression for the radius of the ball of 
accessible states
\begin{equation}\label{eqn:RadiusSphere}
\begin{split}
		r(t)&=e^{-(2n_{\textrm{II}}(\omega_0)+1)J_\textrm{I}(\omega_0) t} \\
		&\exp \left( 2 (n_{\textrm{II}}(\omega_0)-n_{\textrm{I}}(\omega_0)) \frac{J_\textrm{I}(\omega_0)}{J_{\textrm{II}}(\omega_0)} (1-e^{-J_{\textrm{II}}(\omega_0) t})\right)~,
\end{split}
\end{equation}
which in the limit $J_{\textrm{II}}(\omega_0)t \to 0$ becomes
\begin{equation}\label{eqn:RadiusSphereWeak}
	r(t)=e^{-(2n_{\textrm{I}}(\omega_0)+1)J_\textrm{I}(\omega_0) t}~,
\end{equation}
that is the expression that we would obtain if only RI was considered. This means that the rate at which the volume of the accessible states reduces is mainly governed by RI. A smaller coupling between OQS and RI would cause a slower reduction of the accessible states space. 
The center of the ball of accessible states is given by the evolved polarization vector of the maximally mixed state, namely the origin of the Bloch sphere, and is thus at $\vec{c}(t)=(0,0,c(t))$ with
\begin{equation}
	\label{eqn:CenterSphere}
\begin{split}
c(t) &= -J_\textrm{I}(\omega_0) \int_0^t dt' e^{-(2n_{\textrm{II}}(\omega_0)+1)J_\textrm{I}(\omega_0) (t-t')} \\
	\exp &\left( 2 (n_{\textrm{I}}(\omega_0)-n_{\textrm{II}}(\omega_0))  J_\textrm{I}(\omega_0)  \frac{e^{-J_{\textrm{II}}(\omega_0) t}-e^{-J_{\textrm{II}}(\omega_0) t'}}{J_{\textrm{II}}(\omega_0)} \right)~.
\end{split}
\end{equation}

This expression has no analytic solution, but can be solved in the short time limit (ST),
 $J_{\textrm{II}}(\omega_0)t \ll 1$. If the exponentials inside the second factor are Taylor expanded in terms of $J_{\textrm{II}}(\omega_0)t$ and $J_{\textrm{II}}(\omega_0)t' \leq  J_{\textrm{II}}(\omega_0)t$ up to first order, the resulting integral is solvable and yields
\begin{equation}\label{eqn:STcenter}
	c_{\textrm{ST}}(t)=\frac{e^{-J_{\textrm{I}}(2n_{\textrm{I}}(\omega_0)+1)t }-1}{2n_{\textrm{I}}(\omega_0)+1}~.
\end{equation}
Within this regime, we distinguish two limiting cases
\begin{itemize}
	\item When $J_{\textrm{I}}(\omega_0)(2n_{\textrm{I}}(\omega_0)+1)t \ll 1$, Eq.~(\ref{eqn:STcenter}) approximately reduces to 
	\begin{equation}
		c_{\textrm{ST}}(t)\approx\frac{-J_{\textrm{I}}(2n_{\textrm{I}}(\omega_0)+1)}{2n_{\textrm{I}}(\omega_0)+1}t~,
	\end{equation}
	which at time $t=0$ corresponds to the center of the Bloch sphere.
	\item When $J_{\textrm{I}}(\omega_0)(2n_{\textrm{I}}(\omega_0)+1)t \gg 1$, the exponential in Eq.~(\ref{eqn:STcenter}) vanishes, and this expression becomes
	\begin{equation}
		c_{\textrm{ST}}=\frac{-1}{2n_{\textrm{I}}(\omega_0)+1}~,
		\label{pointbloch}
	\end{equation}
	such that $(0,0,c_{\textrm{ST}})$ corresponds to the thermal state $\rho_\textrm{S}^{\textrm{th}}(\beta_\textrm{I})$. 
This expression holds when
\begin{equation}\label{eqn:Restriction}
	J_\textrm{I}(\omega_0)(2 n_\textrm{I}(\omega_0) + 1) \gg J_{\textrm{II}}(\omega_0)~,
\end{equation}
in which case the ball of accessible states is centred around the point corresponding to the thermal state of the OQS at $\beta_\textrm{I}$ as long as $J_{\textrm{II}}(\omega_0)t \ll 1$. Moreover, in this limit the radius of the ball of accessible states Eqs.~(\ref{eqn:RadiusSphere}, \ref{eqn:RadiusSphereWeak}) is close to 0, meaning that the state of the OQS is independent of the initial condition and close to the state $\rho_\textrm{S}^{\textrm{th}}(\beta_\textrm{I})$, which shows that the system thermalizes to $\beta_I$.
\end{itemize}

Eq. \eqref{eqn:Restriction}, shows that the condition for the OQS to prethermalize to $\beta_I$, depends on the relationship of this temperature and the coupling strengths, but is independent of $\beta_{II}$.
In the next section we analyse how $\beta_{\textrm{II}}$ affects the prethermalization.

The long time (LT) limit ($J_{\textrm{II}}(\omega_0)t \to \infty$) of Eq.~(\ref{eqn:CenterSphere}), studied analytically in Appendix \ref{append:CenterAymptotic}, yields
\begin{equation}\label{eqn:CenterAsymptotic}
	c_{\textrm{LT}}=\frac{-1}{2n_{\textrm{II}}(\omega_0)+1}~,
\end{equation}
where the point $(0,0,c_{\textrm{LT}})$ corresponds to the thermal state $\rho_\textrm{S}^{\textrm{th}}(\beta_\textrm{II})$ as the asymptotic state. This asymptotic state was also checked analytically  using the approximate decay rates in Appendix \ref{append:Asymptotic}.

\begin{figure}
\centering
	\includegraphics[trim=7 6.7 0 0,clip,width=\linewidth]{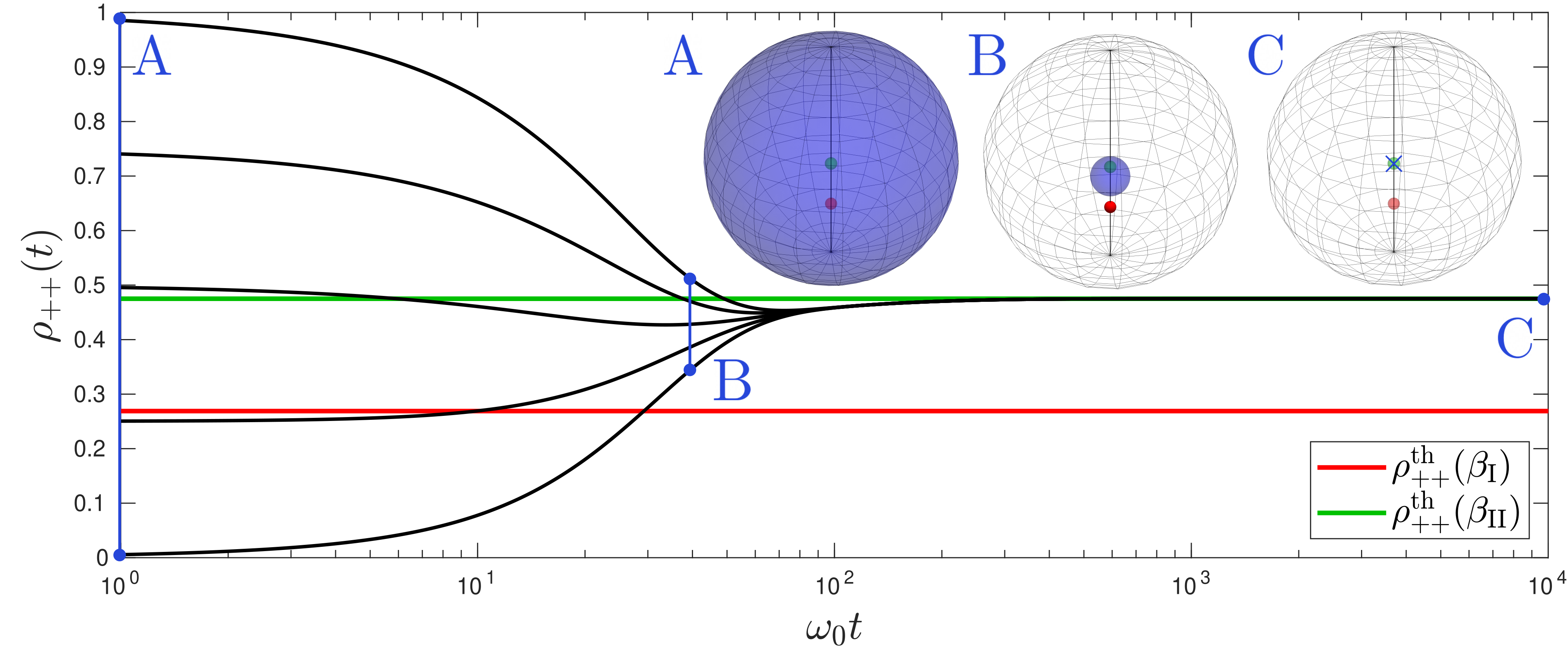}
	\caption{Evolution of the population of the $\ket{+}$ state for different initial conditions of pure states, and snapshots (A,B,C) of the evolution of the ball of accessible states. In this case, in opposition to Fig.~\ref{fig:Pretherm}, all initial condition directly tend to the thermal state at $\beta_{\textrm{II}}$, which is the asymptotic state of the system. The ball of accessible states contracts around the point representing the thermal state at $\beta_{\textrm{II}}$, and after becoming punctual it stays there. There is no prethermalization phenomenon in this case, where the spectral functions parameters are as in Fig.~\ref{fig:Pretherm} with the exception that $g_{\textrm{II}}=10^{-2}$.}
	\label{fig:NoPretherm}
\end{figure}

To illustrate the above discussion, we display in Figs.~\ref{fig:Pretherm} and \ref{fig:NoPretherm} the evolution of the system in two different scenarios. In both cases, the time dependence of the $\rho_{++}(t)=\bra{+}\rho_S(t)\ket{+}$ component\footnote{Where $\ket{\pm}$ is the eigenbasis defined by $H_S$ in Eq.~(\ref{eqn:HamiltonianSystem}) with eigenvalues $\pm \omega_0/2$.} of the state of the system is shown for several initial pure states, which allows us to visualize the evolution of the ball of accessible states.
In the first case, for $\beta_\textrm{I}=1$, $\beta_{\textrm{II}}=0.1$ and $g_{\textrm{II}}=10^{-5}$ we observe prethermalization (Fig.~\ref{fig:Pretherm}), but when the coupling is increased to 
$g_\mathrm{II}=10^{-2}$ (Fig.~\ref{fig:NoPretherm}), the phenomenon does not appear.

Following our previous considerations, we identify two relevant timescales that govern the OQS evolution in the prethermalization regime of Fig. \ref{fig:Pretherm}.
First, the time $t_\textrm{I}$ after which the OQS has evolved to the thermal state at $\beta_\textrm{I}$.
At this time, the space of accessible states has already contracted to a point, so that the state reached is independent of the initial condition. 
The second timescale $t_{\textrm{II}}$ determines the time required for thermalization to the asymptotic state $\rho_\textrm{S}^{\textrm{th}}(\beta_{\textrm{II}})$.
If $t_\textrm{I}$ is sufficiently smaller than $t_\textrm{II}$, as in Fig. \ref{fig:Pretherm}, the system first evolves to $\rho_\textrm{S}^{\textrm{th}}(\beta_\textrm{I})$ (red dot),
and stays close to it for a certain time $t_\textrm{pr}$, which we call prethermalization time.
After this time, it smoothly evolves to $\rho_\textrm{S}^{\textrm{th}}(\beta_{\textrm{II}})$ (green dot).
As shown in Fig. \ref{fig:NoPretherm}, when the conditions of the problem do not allow for prethermalization, we observe the thermalization of any initial condition directly to the state $\rho_\textrm{S}^{\textrm{th}}(\beta_{\textrm{II}})$, without any transitory approach to $\rho_\textrm{S}^{\textrm{th}}(\beta_{\textrm{I}})$.

\begin{figure}
\centering
	\includegraphics[width=\linewidth]{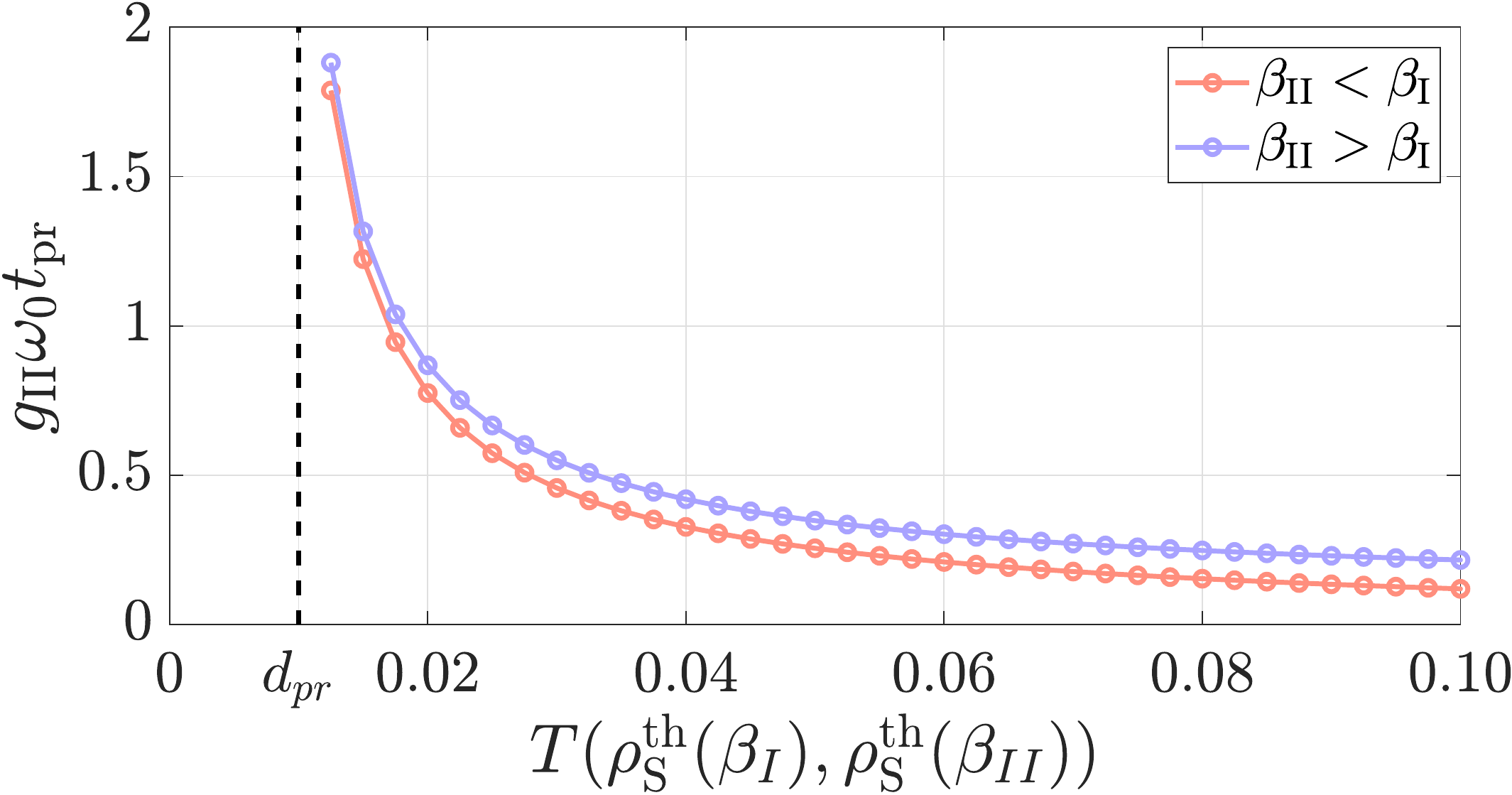}
	\caption{Prethermalization time, as a function of the trace distance between the thermal state of the system at fixed $\beta_\textrm{I}=1.1$ and varying $\beta_\textrm{II}$. When the trace distance is smaller than $d_{\textrm{pr}}=10^{-2}$ prethermalization is not defined. 
	The coupling strength between reservoirs is $g_\textrm{II}=10^{-3}$ and the remaining parameters are the same as in Fig.~\ref{fig:Pretherm}.}
	\label{fig:PrethermTracedist}
\end{figure}

\begin{figure}
\centering
	\includegraphics[width=\linewidth]{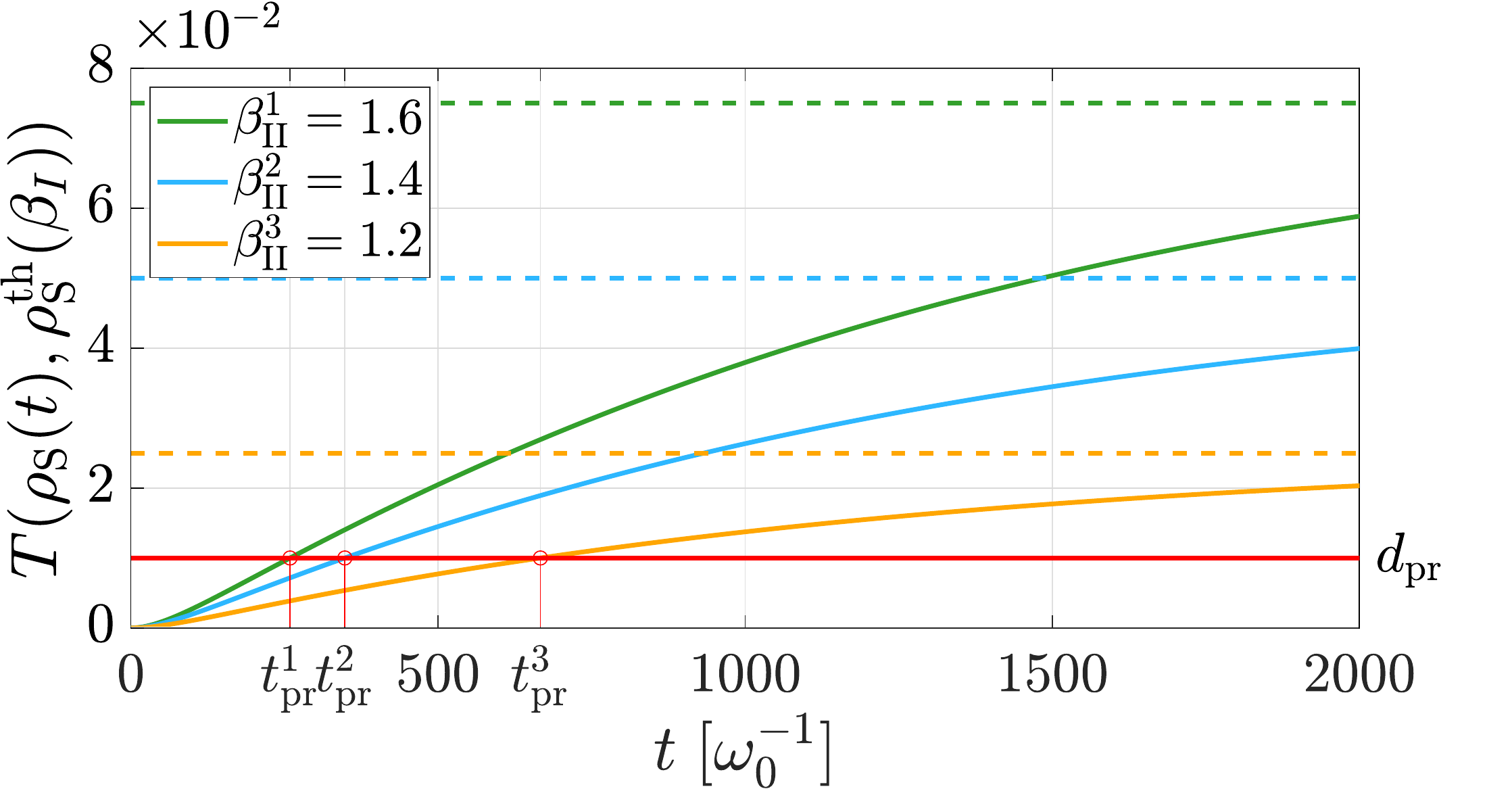}
	\caption{Trace distance between the OQS and its thermal state at a fixed $\beta_\textrm{I}=1.1$ for different values of  $\beta_\textrm{II}^i=\{1.6, 1.4, 1.2\}$. Dashed lines represent the trace distance between thermal states at $\beta_\textrm{I}$ and $\beta_\textrm{II}^i$, i.e., $T(\rho_\textrm{S}^{\textrm{th}}(\beta_{\textrm{I}}),\rho_\textrm{S}^{\textrm{th}}(\beta_{\textrm{II}}^i))$, so that when the OQS approaches the asymptotic state, solid lines tend to the dashed lines.
	 Initially the OQS is in a thermal state at $\beta_\textrm{I}$ and departs from it as it evolves. We observe that this departure happens faster for a larger separation between the thermal states of the environments. The red line represents the distance $d_\textrm{pr}=10^{-2}$ and the rest of the parameters are as in Fig.~\ref{fig:PrethermTracedist}.}
	\label{fig:Scheme}
\end{figure}

\subsection{Prethermalization Time}

To give a more quantitative estimation of the time during which the OQS remains approximately thermalized at the temperature $\beta_\textrm{I}$, i.e. the prethermalization time $t_\textrm{pr}$, we make use of the trace distance
\begin{equation}\label{eqn:TraceDistance}
	T(\rho_1,\rho_2)=\frac{1}{2}\Tr\left\{\sqrt{(\rho_1-\rho_2)^2}\right\}~,
\end{equation}
between the evolved state of the OQS, with initial condition $\rho_\textrm{S}(0)=\rho_\textrm{S}^\textrm{th} (\beta_\textrm{I})$, and the thermal state $\rho_\textrm{S}^{\textrm{th}}(\beta_\textrm{I})$. 
We define $t_\textrm{pr}$ as the time elapsed between the time at which the radius of the ball of accessible states has reduced below 10\%, and the 
time at which the above trace becomes bigger than a fixed trace distance $d_\textrm{pr}$. This represents a threshold distance below which two states could not be distinguished.
If the order in which these events happen is the opposite, it means that no prethermalization is present.

We can visualize this by looking at the dynamics of the polarization vector corresponding to the density matrix of the system, starting from $\rho_\textrm{th}(\beta_\textrm{I})$. If that point has been significantly displaced before the ball of accessible states has contracted, then no prethermalization is present: See Figs.~\ref{fig:Pretherm} and \ref{fig:NoPretherm} for a visual reference of this criterion.  If the trace distance between the thermal state of the system at $\beta_\textrm{I}$ and $\beta_\textrm{II}$ is smaller than $d_{\textrm{pr}}$, the prethermalization time is not defined, as these two states would not be distinguishable.

With this definition we studied how $t_\textrm{pr}$ varies as a function of the initial temperatures of both reservoirs, as well as for different values of the coupling strength between them, i.e. $g_{\textrm{II}}$. 
In Fig.~\ref{fig:PrethermTracedist} we show the prethermalization time as a function of the trace distance for fixed $\beta_{\textrm{I}}$ varying $\beta_{\textrm{II}}$. We observe the prethermalization time to be longer, the closer the two states are. The same can be appreciated in Fig.~\ref{fig:Scheme}, which shows the calculation of $t_{\textrm{pr}}$ for different separations of the thermal states at $\beta_\textrm{I}$ and $\beta_\textrm{II}$: When they are closer (orange line) $t_{\textrm{pr}}$ is higher and when they are further apart (blue and green lines) $t_{\textrm{pr}}$ decreases.

\begin{figure}
\centering
	\includegraphics[width=\linewidth]{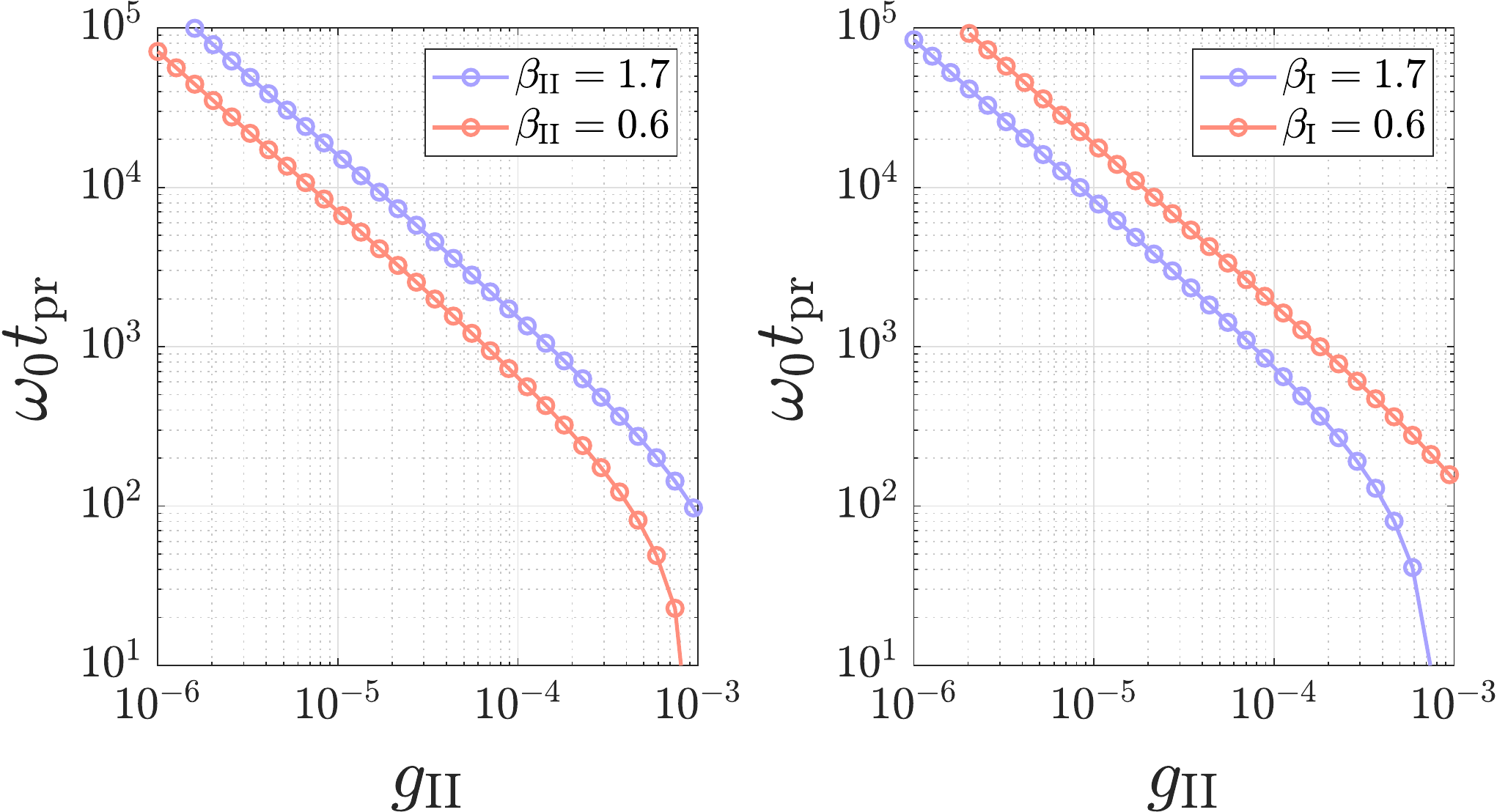}
	\caption{Prethermalization time, as a function of the coupling strength $g_\textrm{II}$, for a fixed $\beta_\textrm{I}=1.1$ varying $\beta_\textrm{II}$ (the left panel) and for fixed $\beta_\textrm{II}=1.1$  varying $\beta_\textrm{I}$ (right panel). The rest of the parameters are the same as in Fig.~\ref{fig:Pretherm}.}
	\label{fig:ThemalizationScale}
\end{figure}

\begin{figure*}
	\centering
	\begin{tabular}{cc}
	
	\begin{TAB}(r,1cm,4cm){c}{cc}
	\begin{tabular}{l}
	\Large{(a)}\\
	
		\\
		\begin{tikzpicture}[scale=0.8]
		\tikzstyle{S}=[draw,circle,fill=blue!50!white,text=black,minimum width=20pt]
		\tikzstyle{RI}=[draw,circle,fill=red!50!white,text=black,minimum width=25pt]
		\tikzstyle{RII}=[draw,circle,fill=green!30!white,text=black,minimum width=50pt,align=left]
		\tikzstyle{RIIname}=[draw,circle,color=green!30!white,fill=green!30!white,text=black,minimum width=10pt,align=left]
		  \foreach \place/\name in {{(0,0)/S}}
		    \node[S] (\name) at \place {\Large \name};
		        
		  \foreach \place/\name/\cosa in {{(1.55,0)/RIR/$\textrm{RI}^\textrm{R}$},{(-1.55,0)/RIL/$\textrm{RI}^\textrm{L}$}}
		    \node[RI] (\name) at \place{\cosa};
		  
		  \foreach \source/\dest in {S/RIR, S/RIL}
		    \path(\source) edge (\dest);
		\end{tikzpicture}\\
		 \\
		 \\
		\end{tabular}\\
		\begin{tabular}{l}
	\Large{(b)}\\
		 \\
		\begin{tikzpicture}
		\tikzstyle{S}=[draw,circle,fill=blue!50!white,text=black,minimum width=20pt]
		\tikzstyle{RI}=[draw,circle,fill=red!50!white,text=black,minimum width=25pt]
		\tikzstyle{RII}=[draw,circle,fill=green!30!white,text=black,minimum width=53pt,align=left]
		\tikzstyle{RIIname}=[draw,circle,color=green!30!white,fill=green!30!white,text=black,minimum width=10pt,align=left]
		  \foreach \place/\name in {{(0,0)/S}}
		    \node[S] (\name) at \place {\Large \name};
		    
		  \foreach \place/\name/\cosa in {{(1.5,0)/RIIR/RII},{(-1.5,0)/RIIL/RII}}
		    \node[RII] (\name) at \place{ };    
		    
		   \foreach \place/\name/\cosa in {{(1.9,0)/RIIR/$\textrm{RII}^\textrm{R}$},{(-1.9,0)/RIIL/$\textrm{RII}^\textrm{L}$}}
		    \node[RIIname] (\name) at \place{ \cosa};    
		    
		  \foreach \place/\name/\cosa in {{(1.02,0)/RIR/$\textrm{RI}^\textrm{R}$},{(-1.02,0)/RIL/$\textrm{RI}^\textrm{L}$}}
		    \node[RI] (\name) at \place{\cosa};
		  
		  \foreach \source/\dest in {S/RIR, S/RIL}
		    \path(\source) edge (\dest);
		\end{tikzpicture}
		\\
		 \\
		 \\
		\end{tabular}
\end{TAB}
&\includegraphics[width=0.71\linewidth]{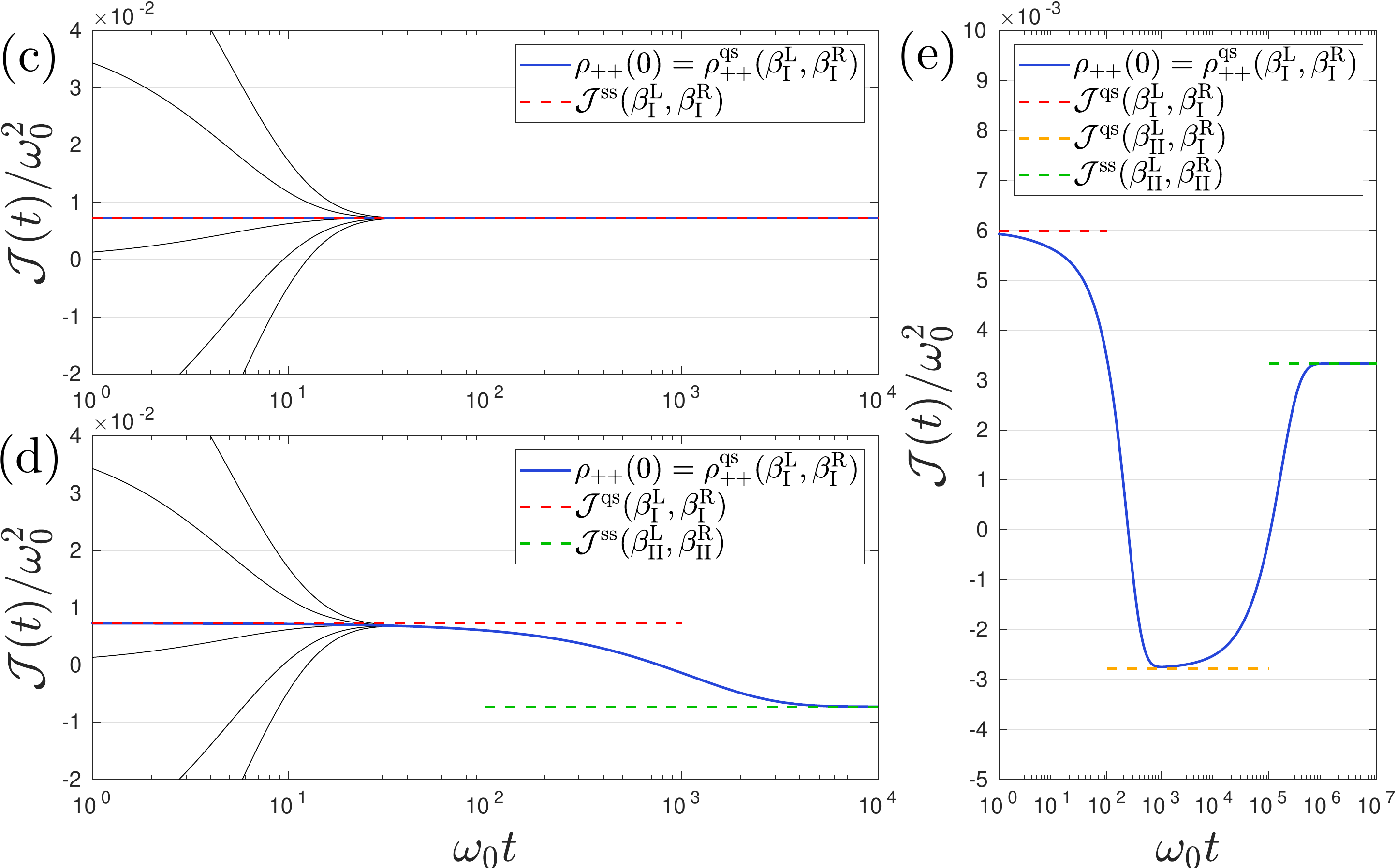}\\
	\end{tabular}
	\caption{Panel (c) represents the heat flux corresponding to the system coupled to two equilibrium environments, as depicted in panel (a). Panel (d) represents the heat flux where each of the environments are out of equilibrium, as depicted in panel  (b). Both panels (c) and (d)  show how different initial conditions evolve, after some time, to a flux $\mathcal{J}^{\textrm{ss(qs)}}\left(\beta_\textrm{I}^\textrm{L},\beta_\textrm{I}^\textrm{R}\right)$. However, while in the simple case (c) this flux remains constant, in (d) the system continues to evolve to a final flux given by $\beta_{\textrm{II}}^\textrm{L} $ and $\beta_{\textrm{II}}^\textrm{R}$.
	The parameters are $\omega_0=1$, $g_\textrm{I}=10^{-2}$, $g_{\textrm{II}}=10^{-3}$, $s_\textrm{I}=s_{\textrm{II}}=1$, $\omega_{c\textrm{I}}=\omega_{c\textrm{II}}=10$ for both environments, while the temperatures differ as $\beta_{\textrm{I}}^\textrm{L} = 1$, $\beta_{\textrm{I}}^\textrm{R} = 0.1$, $\beta_{\textrm{II}}^\textrm{L} = 0.1$ and $\beta_{\textrm{II}}^\textrm{R} = 1$. We plotted a special initial condition in blue that starts in the steady state defined by the temperatures of the first reservoirs. Panel (e) shows a more complex evolution of the fluxes in which there are two sign flips of the heat flux. The parameters of this simulation are  $\beta_{\textrm{I}}^\textrm{R} = 0.1$, $\beta_{\textrm{I}}^\textrm{L} = 0.5$, $\beta_{\textrm{II}}^\textrm{R} = 1$, $\beta_{\textrm{II}}^\textrm{L} = 10$, $g_{\textrm{II}}^\textrm{R}=10^{-2}$ and $g_{\textrm{II}}^\textrm{L}=10^{-5}$, while the rest of parameters are the same as the other plots.}
	\label{fig:HeatFluxes}
\end{figure*}

We realized that the prethermalization scale for fixed temperatures of both reservoirs depends inversely with  $g_{\textrm{II}}$, i.e., $t_{\textrm{pr}} \sim g_\textrm{II}^{-1}$, for small values of $g_\textrm{II}$. 
This can be clearly seen, for small coupling strenghts, in Fig.~\ref{fig:ThemalizationScale}.
 In this figure we checked that condition \eqref{eqn:Restriction} is fulfilled for all points.  It was pointed out in \cite{Rigol-PhysRevX.9.021027} that the thermalization time $t_\textrm{II}$ scales as $g_\textrm{II}^{-1}$ similarly as $t_\textrm{pr}$, and we checked that this is true for our case\footnote{In \cite{Rigol-PhysRevX.9.021027} the thermalization timescales with $g^{-2}$, where the weakly perturbed Hamiltonian is proportional to $g$. In our case the perturbation is proportional to $g_{\lambda,k}$ and $g_\textrm{II} \propto |g_{\lambda,k}|^2$.} as well.

From Figs.~\ref{fig:PrethermTracedist} and \ref{fig:ThemalizationScale} we also observe that the time the OQS stays in the prethermal state is shorter for a higher temperature of RII (lower $\beta_{\textrm{II}}$), while the duration of the prethermalized state increases with $\beta_{\textrm{II}}$. A similar qualitative behaviour is observed for varying $\beta_{\mathrm{I}}$ (see right panel of Fig. \ref{fig:ThemalizationScale}), but the overall prethermalization scale is smaller in case of a larger $\beta_{\mathrm{I}}$.

\section{Composite non-equilibrium Environments}
\label{VI}

In the scenario discussed in the previous sections, the OQS is expected to reach a steady state at thermal equilibrium, consistent with the temperature of the largest reservoir RII.
However, using the same ME formalism it is also possible to construct more complex scenarios in which the steady state of the open system is out of equilibrium, for instance when the system is coupled to 
 two independent environments, as depicted in the left column of Fig.~\ref{fig:HeatFluxes}. 
The additional environment gives rise to a new dissipative term in the master equation (\ref{eqn:CanonicalME}) and a new term in the environment corrected Hamiltonian $H(t)$. The corresponding rates and Lamb shift corrections are computed identically as before. 
The heat flow between environments produces a change in the energy of the system 
$E_\textrm{S}=\text{Tr}\{H_\textrm{S} \rho_\textrm{S}(t)\}$, where $H_S$ is the system Hamiltonian. The evolution of this quantity can be expressed by the canonical ME (\ref{eqn:CanonicalME}) as
\begin{equation}
\begin{split}
	\frac{d E_\textrm{S}}{dt}  = -&i \Tr \left\{ H_\textrm{S}  [\tilde{H}(t),\rho_\textrm{S}(t)] \right\}\\
	 + &\sum_{\nu=\textrm{L},\textrm{R}}   \Tr \left\{  H_\textrm{S} \mathcal{D}^{(\nu)}(t,\rho_\textrm{S}(t)) \right\} ~,
\end{split}
\end{equation}
where the superindex $(\nu)=\{\textrm{L},\textrm{R}\}$ refers to the left and right environments and $\tilde{H}(t)=H_\textrm{S} + \sum_\nu \frac{1}{2}\Delta\omega^{(\nu)}(t)\sigma_z$. Since $\tilde{H}(t)$ and $H_\textrm{S}$ are both proportional to $\sigma_z$, the first term vanishes, and the second one defines the heat fluxes
\begin{equation}\label{eqn:HeatFluxDef}
	\mathcal{J}^{(\nu)}(t)=\Tr \left\{  H_\textrm{S} \mathcal{D}^{(\nu)}(t,\rho_S(t)) \right\}~,
\end{equation}
from environment $(\nu)$ to the OQS. By convention, we consider the heat flux from the right reservoir to be positive, and the one from the left to be negative. Thus, a positive total heat flux indicates a flow of energy from right to left, and vice versa. In the following subsections we first analyze the heat fluxes when the left and right reservoirs are each in a thermal equilibrium state, and then when each of them are out of equilibrium.

\subsection{Heat flux between environments in equilibrium}

In the case of equilibrium environments of Fig.~\ref{fig:HeatFluxes}a, which we depict as single reservoirs on each side of the OQS, any initial state reaches a non equilibrium steady state that depends on the initial state of both environments under our model assumptions.
In Fig.~\ref{fig:HeatFluxes}c we plot the total heat flux $\mathcal{J}^\textrm{L}(t)+\mathcal{J}^\textrm{R}(t)$ calculated using Eq.~(\ref{eqn:HeatFluxDef}) with $\gamma_+^{(\nu)}=J_\textrm{I}^{(\nu)}(\omega_0) n_\textrm{I}^{(\nu)}(\omega_0)$ and $\gamma_-^{(\nu)}=J_\textrm{I}^{(\nu)}(\omega_0) (n_\textrm{I}^{(\nu)}(\omega_0)+1)$ the decay rates of the spin boson model with one reservoir. 
We observe that initially, the heat flux depends on the initial condition, but after some time it always converges to the value in the steady state,
\begin{equation}
	\mathcal{J}^{\textrm{ss}(\nu)}=\omega_0 \left[ \gamma_+^{(\nu)}- \rho_{++}^{\textrm{qs}}(\beta_\textrm{I}^\textrm{L},\beta_\textrm{I}^\textrm{R}) \left( \gamma_+^{(\nu)}+\gamma_-^{(\nu)}\right) \right]~,
\end{equation}
where $\rho_{++}^{\textrm{qs}}(\beta_\textrm{I}^\textrm{L},\beta_\textrm{I}^\textrm{R})$ is given in Appendix \ref{append:NonEqAsymptotic}. 
When the OQS reaches the asymptotic state there is a constant heat flux from the environment with the higher temperature. In Fig.~\ref{fig:HeatFluxes}c, which corresponds to $\beta_\textrm{I}^\textrm{R}<\beta_\textrm{I}^\textrm{L}$, this is observed by a positive steady state flux.

\subsection{Heat flux between environments that are out of equilibrium}

We now consider the case where the OQS is coupled to two out of equilibrium reservoirs, as schematically depicted in Fig.~\ref{fig:HeatFluxes}b. Fig.~\ref{fig:HeatFluxes}d shows that the heat flux, independently of the initial condition, is dominated by the temperature gradient between $\beta_\textrm{I}^R$ and $\beta_\textrm{I}^L$,  while the gradient for $\beta_{\textrm{II}}^{(R,L)}$ becomes relevant at longer times. The timescale in which each gradient is dominant is determined by $g_\textrm{II}^{(\nu)}$. Interestingly, we observe that the interplay between these gradients may even produce a change of sign in the current. This is because we have chosen $\beta_\textrm{I}^\textrm{R}<\beta_\textrm{I}^\textrm{L}$, but $\beta_{\textrm{II}}^\textrm{R}>\beta_{\textrm{II}}^\textrm{R}$, such that the quasi-stationary flux is positive (the right RI is hotter than the left RI), while at long times is negative (since the right RII is colder than the left RII).  

Moreover, one can tune these gradients and the couplings $g_\textrm{II}^{(\nu)}$ to be such that there are two changes of sign in the heat current. This is observed in Fig.~\ref{fig:HeatFluxes}e, where $\beta_\textrm{I}^\textrm{R} < \beta_\textrm{I}^\textrm{L} < \beta_\textrm{II}^\textrm{R} < \beta_\textrm{II}^\textrm{L}$ and $g_\textrm{II}^\textrm{R} > g_\textrm{II}^\textrm{L}$, which leads to an initial and final positive flux ($\beta_i^\textrm{R}<\beta_i^\textrm{L}$). But as $g_\textrm{II}^\textrm{R} > g_\textrm{II}^\textrm{L}$ there is some time that the quasi-stationary flux is determined by $\beta_\textrm{I}^\textrm{L} < \beta_\textrm{II}^\textrm{R}$, such that the flux during that time is negative.

The stationary and quasi-stationary states for the setup Fig.~\ref{fig:HeatFluxes}b can be explicitly derived, and are shown in  Appendix~\ref{append:NonEqAsymptotic}.

\section{Conclusions}
\label{VII}

We have presented a model to describe an OQS which is coupled to a hierarchy of environments at different temperatures, a situation that can be found in complex environments and interfaces that are present in both natural and quantum technological scenarios. 
Although these situations are in principle very complex to analyse, we have shown here that, under certain constraints, one can extract a well-behaved master equation that allows such a description in relevant limits. 

In detail, we have considered an
open system directly coupled to a reservoir RI, at an inverse temperature $\beta_I$, that is driven out of equilibrium because of its coupling to a second reservoir RII at $\beta_{II}$. With the use of weak coupling and Markovian approximations, we have derived a master equation to describe the evolution of the reduced density matrix of the system, by tracing out the evolution of the environment. Even with these approximations, we were able to observe a rich dynamics of the open system, with the existence of a transitory state, called prethermal state, before the final thermalization, which was found to be determined by the larger reservoir solely.  
We investigated under which conditions prethermalization is present, and concluded that this state is longer lived when the reservoir RI, directly coupled to the OQS, is hotter and RII colder, as well as when the coupling between reservoirs is the smallest possible. We presented a way to characterize prethermalization that is independent of the initial condition of the OQS, through the evolution of the volume of accessible states.

We have also shown that non-trivial dynamics and competing timescales are also present when we consider two out of equilibrium environments coupled to the system. It is well-known that, in the standard situation where the environments are in equilibrium, a heat flux with a given direction (from the hot to the cold reservoir) is established and prevails at long times. 
Interestingly, when considering out of equilibrium environments we observe that the timescales induced by different environments may induce that the heat flux switches direction, even more than once.

 As shown, the OQS dynamics and its currents do not evolve according to a single timescale, but present a richer dynamics that may be evident in experiments and quantum information processes,  particularly at long times. The presence of a prethermalization transitory may be harnessed in quantum technological applications, for instance by considering the initialization protocols of a qubit based on coupling it to a reservoir \cite{Tuorila2017,PhysRevResearch.1.013004}. The added reservoir can potentially be controlled by a second one, according to our scheme, in order to optimize further the protocol. In other words, our work describes the possibility of  manipulating and controlling an open system by externally modifying and controlling the reservoir to which it is directly coupled.

Our scheme can be adapted to include more external reservoirs at different temperatures. Multiple layer environments can be found, for instance, in superconducting quantum computers, where qubits are affected not only by surrounding layers cryogenically cooled, but also by outer layers at increasingly higher temperatures. Considering this reservoir structure would allow us to find additional transitory and steady states of the OQS, which can potentially be harnessed and controlled.
An interesting subject for further investigation would also be the consideration of the dynamics beyond the weak-coupling approximation, and the inclusion of non-Markovian effects.

\acknowledgments
This work has been partly funded by the  CNRS PEPs Spain-France PIC2017FR6, the Spanish FEDER/MCIyU-AEI grant FPA2017-84543-P, SEV-2014-0398, Generalitat Valenciana grant PROMETEO/2019/087
and the Deutsche Forschungsgemeinschaft (DFG, German Research Foundation) under Germany's Excellence Strategy -- EXC-2111 -- 390814868. We also acknowledge support from CSIC Research Platform PTI-001. IDV acknowledges finantial support by DFG-grant GZ: VE 993/1-1.

\appendix 

\section{Canonical Master Equation, decay rates and frequency shift}\label{append:ME}

For the interaction Hamiltonian considered in Eq.~(\ref{eqn:HamInter}),
the canonical decay rates and decoherence channels of the master equation
(\ref{eqn:ME}) are 
\begin{align}
\gamma_{1}(t) & =P(t)+P^{*}(t) \equiv \gamma_+(t) ~, & L_{1}(t) & =\sigma^+ \equiv L_+~,\nonumber \\
\gamma_{2}(t) & =M(t)+M^{*}(t) \equiv \gamma_-(t)~, & L_{2}(t) & =\sigma^- \equiv L_-~,\nonumber \\
\gamma_{3}(t) & =0~, & L_{3}(t) & =\frac{1}{\sqrt{2}}\sigma_{z}~,\numberthis\label{eqn:CanonicalDecay}
\end{align}
where we defined 
\begin{equation}\label{eqn:PlusCorrel}
P(t)=\int_{0}^{t}dt'\alpha^{+}(t,t')e^{-i\omega_{0}t'}~,
\end{equation}
and 
\begin{equation}\label{eqn:MinusCorrel}
M(t)=\int_{0}^{t}dt'\alpha^{-}(t,t')e^{i\omega_{0}t'}~.
\end{equation}
The operator $H(t)$ is a modification of the free Hamiltonian
of the OQS 
\begin{equation}\label{eqn:ShiftHam}
H(t)=H_{\textrm{S}}+\frac{1}{2}\Delta\omega(t)\sigma_{z}~,
\end{equation}
which, in this case, represents a shift of the natural frequency of the system, given
by 
\begin{equation}\label{eqn:ShiftImaginaryPart}
\Delta\omega(t)=\frac{i}{2}(P(t)-P^{*}(t))-\frac{i}{2}(M(t)-M^{*}(t))~.
\end{equation}
Therefore, this Hamiltonian can be rewritten as 
\begin{equation}
H(t)=\frac{1}{2}\Omega(t)\sigma_{z}~,
\end{equation}
where $\Omega(t)=\omega_{0}+\Delta\omega(t)$, is the shifted
frequency of the OQS due to the action of the environment. 
The ME for the different matrix elements of the reduced density matrix reads
\begin{equation}\label{eqn:MEmatrix}
\begin{split}
	\dot{\rho}_{++}(t)=& \gamma_+(t) - \rho_{++}(t) [\gamma_+(t)+\gamma_-(t)]~, \\
	\dot{\rho}_{+-}(t) = &\left\{-i \Omega(t) - [\gamma_+(t)+\gamma_-(t)] \right\}\rho_{+-}(t) ~,
\end{split}
\end{equation}
where $\rho_{++}(t)=\langle + |\rho_\textrm{S}(t)|+\rangle$ is the upper population and $\rho_{+-}(t)=\langle + |\rho_\textrm{S}(t)|-\rangle$ is the coherence, in the $\ket{\pm}$ eigenbasis of $H_S$. We made use of the trace preservation of the dynamical map.

\section{Evolution of RI operators}\label{append:OperatorA}

The time evolution of the operator $a_{\lambda}(t)$
is given by the Heisenberg equation 
\begin{equation}
\frac{d}{dt}a_{\lambda}(t)=i\left[H_{\textrm{E}},a_{\lambda}(t)\right]=-i\omega_{\lambda}a_{\lambda}(t)-i\sum_{k}\tilde{g}_{\lambda k}b_{\lambda k}(t)~,\label{eqn:HeisenA}
\end{equation}
where $b_{\lambda,k}(t)$ is, in turn, given by its corresponding
equation 
\begin{equation}
\frac{d}{dt}b_{\lambda k}(t)=i\left[H_{\textrm{E}},b_{\lambda k}(t)\right]=-i\omega_{\lambda,k}b_{\lambda k}(t)-i\tilde{g}_{\lambda k}a_{\lambda}(t)~.\label{eqn:HeisenB}
\end{equation}
Formal integration of the latter and substitution on the former
yields 
\begin{equation}\label{eqn:OperatorAInteraction}
\begin{split}\frac{d}{dt}\tilde{a}_{\lambda}(t)= & -i\sum_{k}\tilde{g}_{\lambda k}b_{\lambda k}(0)e^{-i(\omega_{\lambda,k}-\omega_{\lambda})t}\\
 & -\sum_{k}\tilde{g}_{\lambda k}^{2}\int_{0}^{t}dt'e^{-i(\omega_{\lambda,k}-\omega_{\lambda})(t-t')}\tilde{a}_{\lambda}(t')~,
\end{split}
\end{equation}
where we also performed the change of variable $\tilde{a}_{\lambda}(t)=e^{i\omega_{\lambda}t}a_{\lambda}(t)$ in order to separate the free evolution part of this operator for a better implementation of the following approximation. 
The first term on the r.h.s. is the quantum noise originated by RII. The
second term can be simplified under the Weisskopf-Wigner approximation
\cite{Weisskopf1997,scully_zubairy_1997}, where the operator $\tilde{a}_{\lambda}(t)$
is assumed to vary with a rate slower than $\omega_{\lambda}$. This
allows us to move the operator $\tilde{a}_{\lambda}(t)$ outside the
integral and, since the exponential inside the integral evolves faster
than $\tilde{a}_{\lambda}(t)$, to extend the integration limit to infinity,
i.e. 
\begin{equation}\label{eqn:MarkovApprox}
\int_{0}^{t}dt'e^{i(\omega_{\lambda}-\omega_{\lambda,k})(t-t')}\tilde{a}_{\lambda}(t')\approx
  \tilde{a}_{\lambda}(t)\int_{0}^{\infty}d\tau e^{i(\omega_{\lambda}-\omega_{\lambda,k})\tau}~,
\end{equation}
where the change of variable $\tau=t-t'$ has been performed. The
above approximation is, in fact, a Markovian approximation for the interaction with RII, since
the operator $\tilde{a}_{\lambda}(t)$ only depends on $t$, so that we
have neglected its past evolution. This integral can be solved via the Sokhotski-Plemelj theorem
by rewriting the second term in the r.h.s. of Eq.~(\ref{eqn:OperatorAInteraction}) as $\gamma_{\lambda}\tilde{a}_{\lambda}(t)$,
where we defined the damping constant 
\begin{equation}\label{eqn:decayrate}
\gamma_{\lambda}=\pi \sum_k|\tilde{g}_{\lambda k}|^{2}\delta(\omega_{\lambda,k}-\omega_\lambda) -i\sum_{k}|\tilde{g}_{\lambda k}|^{2}\mathcal{P}\left(\frac{1}{\omega_{\lambda,k}-\omega_{\lambda}}\right)~.
\end{equation}
This approximation
allows for an exact solution of Eq.~(\ref{eqn:OperatorAInteraction}),
which after undoing the change of variable introduced above leads to 
\begin{equation}
a_{\lambda}(t)=a_{\lambda}(0)e^{-(i\omega_{\lambda}+\gamma_{\lambda})t}+\int_{0}^{t}dt'e^{-(i\omega_{\lambda}+\gamma_{\lambda})(t-t')}f_{\lambda}(t')~,\label{eqn:SolOperatorA}
\end{equation}
where we introduced 
\begin{equation}
f_{\lambda}(t)=-i\sum_{k}\tilde{g}_{\lambda k}b_{\lambda k}(0)e^{-i\omega_{\lambda,k}t}~.
\end{equation}

\section{Correlation function of the ME}\label{append:correlME}
Once the time dependence of the operator $B(t)=\sum_\lambda g_ \lambda a_\lambda(t)$ is explicitly known, we can compute the correlation functions (\ref{eqn:CorrelDef}). First notice that the first (second) term of Eq.~(\ref{eqn:SolOperatorA}) is lineal in operators acting on RI (RII), so that the traces, which are linear in these operators, will be null, and only the quadratic ones will yield non vanishing terms. In this way $\alpha^+(t,\tau)$ becomes
\begin{equation}\label{alphaplusmodelB}
\begin{split}
	\alpha^+(t,\tau)=\sum_{\lambda,\lambda'} g_\lambda^* g_{\lambda'} &e^{(i \omega_\lambda - \gamma_\lambda)t} e^{(-i \omega_{\lambda'} - \gamma_{\lambda'})\tau} T^\textrm{I}_{\lambda,\lambda'} \\
	+ \sum_{\lambda,\lambda'} g_\lambda^* g_{\lambda'}&\int_0^t dt' \int_0^\tau dt''  T^{\textrm{II}}_{\lambda,\lambda'}(t',t')\\
	&e^{(i \omega_\lambda - \gamma_\lambda)(t-t')} e^{(-i \omega_{\lambda'} - \gamma_{\lambda'})(\tau-t'')} ,
\end{split}
\end{equation}
where the first trace is
\begin{equation}
T^\textrm{I}_{\lambda,\lambda'}=\text{Tr}_{\textrm{I}}\{a_\lambda^\dagger a_{\lambda'} \rho_{\textrm{I}}(0)\}  \cancelto{1}{\text{Tr}_{II}\{\rho_{II}(0)\}}=\delta_{\lambda,\lambda'} n_\textrm{I}(\omega_\lambda)~,
\end{equation}
and similarly the second one gives
\begin{equation}
\begin{split}
 T^{\textrm{II}}_{\lambda,\lambda'}(t',t')=\text{Tr}_{\textrm{I},\textrm{II}}\{f_\lambda^\dagger(t') f_{\lambda'}(t'') (\rho_{\textrm{I}}(0)\otimes \rho_{\textrm{II}}(0))\}\\
=\sum_{k,k'} \tilde g_{\lambda k}^* \tilde g_{\lambda' k'} e^{i \omega_{\lambda,k}t'}e^{-i \omega_{\lambda',k'} t''} n_{\textrm{II}}(\omega_{\lambda k}) \delta_{\lambda,\lambda'}\delta_{k,k'}~,
\end{split}
\end{equation}
where the commutation relations (\ref{eqn:CommutationRelations}) have been used, and $n_i (\omega)=[\exp(\beta_i\omega) - 1]^{-1}$ is the average thermal number of quanta in the mode $\omega$ at an inverse temperature $\beta_i$, for the $i=\{\textrm{I},\textrm{II}\}$ environment. After these results, it remains to perform the sums in $\lambda'$ and $k'$, and the integrals of the second term to yield
\begin{equation}\label{eqn:alphaplus}
\begin{split}
	\alpha^+&(t,\tau)=\sum_\lambda |g_\lambda|^2 n_\textrm{I}(\omega_\lambda) e^{i \omega_\lambda (t-\tau)}  e^{-\gamma_\lambda (t+\tau)}\\
	 &+ \sum_{\lambda,k} |g_\lambda|^2 |\tilde g_{\lambda k}|^2 n_{\textrm{II}}(\omega_{\lambda k}) C_{\lambda,k}(t,\tau)~,
\end{split}
\end{equation}
where we defined
\begin{equation}
C_{\lambda,k}(t,\tau)=\frac{e^{i \omega_{\lambda,k}t}  - e^{(i\omega_\lambda - \gamma_\lambda)t}}{-i (\omega_\lambda-\omega_{\lambda,k}) + \gamma_\lambda}\frac{e^{-i \omega_{\lambda,k}\tau}  - e^{(-i\omega_\lambda - \gamma_\lambda)\tau}}{i (\omega_\lambda-\omega_{\lambda,k}) + \gamma_\lambda}~.
\end{equation}
By following similar steps one arrives, for $\alpha^-(t,\tau)$, to
\begin{equation}\label{eqn:alphaminus}
\begin{split}
	\alpha^-&(t,\tau)=\sum_\lambda |g_\lambda|^2 (n_\textrm{I}(\omega_\lambda)+1) e^{-i \omega_\lambda (t-\tau)}  e^{-\gamma_\lambda (t+\tau)}\\
	 &+ \sum_{\lambda,k} |g_\lambda|^2 |\tilde g_{\lambda k}|^2 (n_\textrm{II}(\omega_{\lambda k})+1) C_{\lambda,k}^*(t,\tau)~.
\end{split}
\end{equation}

The spectral function is related to the couplings in Eqs.~(\ref{eqn:alphaplus},\ref{eqn:alphaminus}) in the following way
\begin{equation}\label{eqn:DefSpectral}
	J_\textrm{I}(\omega)=2 \pi \sum_\lambda |g_\lambda|^2 \delta(\omega_\lambda-\omega)~,
\end{equation}
for RI and 
\begin{equation}\label{eqn:DefSpectral2}
	J_{\textrm{II}}^\lambda(\omega)=2 \pi \sum_k |\tilde g_{\lambda,k}|^2 \delta(\omega_{\lambda,k}-\omega)~,
\end{equation}
for RII, where the $\lambda$ index in $J_{\textrm{II}}^\lambda(\omega)$ corresponds to the reservoir to which mode $a_\lambda$ is coupled to. We will assume that all the reservoirs that surround any $a_\lambda$ are identical, so that we drop the $\lambda$ dependence on the spectral function of RII. These definitions allow us to reformulate the problem in integral form.

Since in the master equation (\ref{eqn:ME}) the system operators evolve with $\tau-t$, it is suitable to introduce the change of variable $t'=t-\tau$. 
With these considerations one obtains the correlation functions in Eq.~(\ref{eqn:CorrelationFunctions}) in integral form. We also comment that we have neglected the imaginary part of (\ref{eqn:decayrate}) such that $\gamma_\lambda=J_{\textrm{II}}(\omega_\lambda)/2$. It is well-known that the contribution of the imaginary part of Eq.~(\ref{eqn:decayrate}) can be re-casted as a Lamb shift Hamiltonian of the form $H_{\textrm{RI}}^{\textrm{LS}}=\sum_\lambda \gamma^{\textrm{imag}}_\lambda a^\dagger_\lambda a_\lambda$, where $\gamma_\lambda^{\textrm{imag}}=\Im\{\gamma_\lambda\}$. This Hamiltonian is diagonal with $H_{\textrm{RI}}$, and therefore only contributes as a shift to the energies $\omega_\lambda$ that is not relevant for our analysis.

\section{Validity of approximate decay rates}\label{append:ApproxDecay}

\begin{figure}
\centering
	\includegraphics[width=\linewidth]{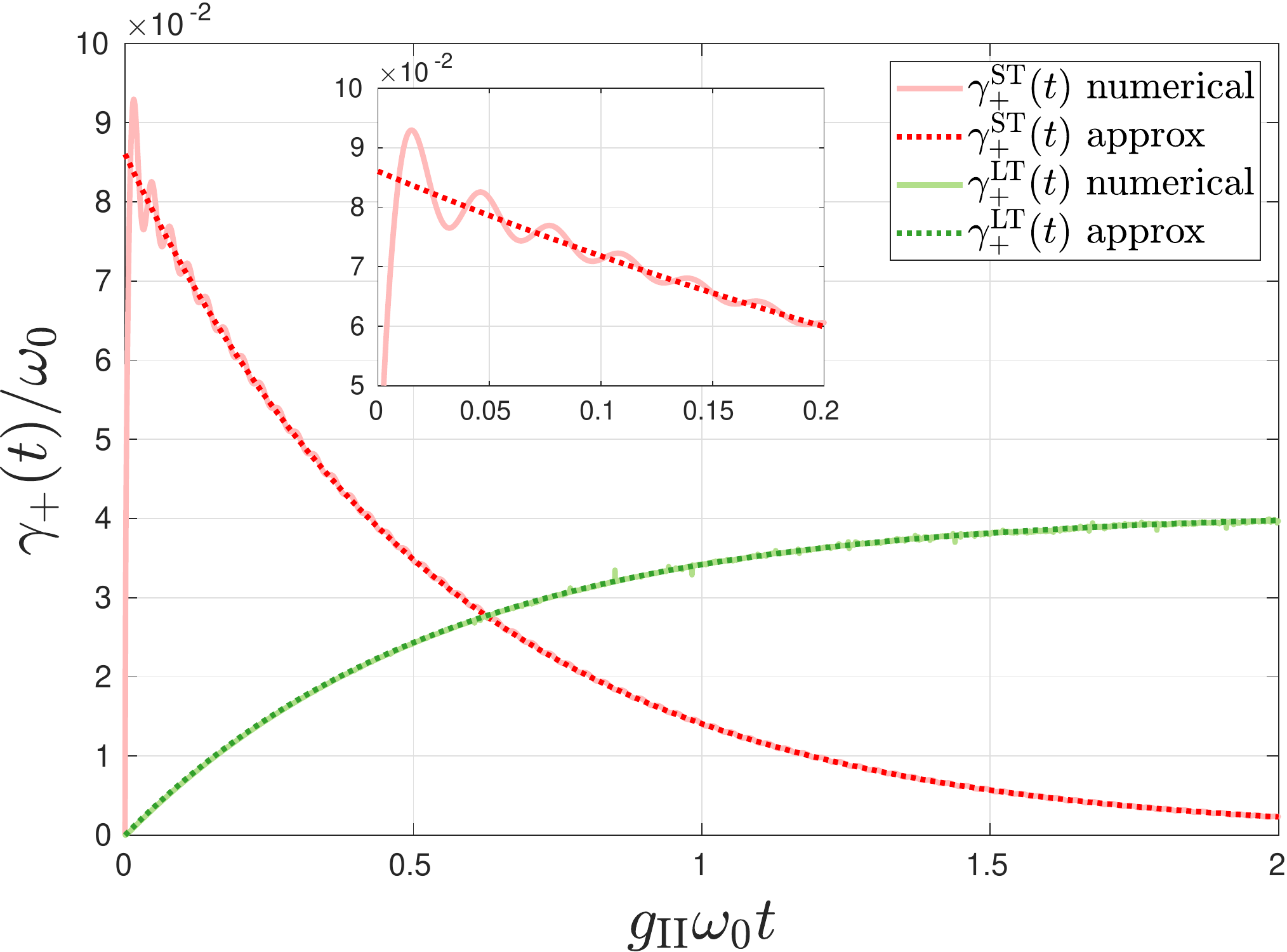}
	\caption{Comparison between the exact and the approximation of each term of the canonical decay rates. The environments parameters are $\omega_0=1$, $g_\textrm{I}=g_{\textrm{II}}=10^{-2}$, $s_\textrm{I}=s_{\textrm{II}}=1$, $\omega_{c\textrm{I}}=\omega_{c\textrm{II}}=10$, $\beta_\textrm{I}=0.1$ and $\beta_{\textrm{II}}=0.2$. The long term exact solution presents some numerical noise.}
	\label{fig:TestApprox}
\end{figure}

The validity of the approximation of the Lorentzian kernel $K(\omega,\omega')$ in Eq.~(\ref{eqn:Kernel}) by a delta function in order to obtain an analytical expression for the canonical decay rates, depends on the parameters of the spectral functions (\ref{eqn:SpectralDens}) for both environments. We have numerically checked the accuracy of the approximation for a broad range of parameters that is relevant for our study. This is illustrated in Fig.~\ref{fig:TestApprox}, where we compare both terms of $\gamma_+(t)$ with the exact result (numerical integration of Eq. (\ref{eqn:PlusCorrel})). The ST term does not reproduce the oscillations of the exact solution, while the LT term perfectly matches the exact result. 

We also considered a more accurate approximation, which consists in taking $J_\mathrm{II}(\omega)$ independent of $\omega$, instead of approximating $K(\omega,\omega_0)$
 by a delta function. This allowed to resolve the oscillatory nature of $\gamma_{\pm}^{\textrm{ST}}$, which is of frequency $\omega_0$, but the result is not as intuitive as the decay rates (\ref{eqn:ApproxDecayRates1},\ref{eqn:ApproxDecayRates2}).

\section{Equilibrium Asymptotic State}\label{append:Asymptotic}

With the help of the approximation (\ref{eqn:ApproxDecayRates1}, \ref{eqn:ApproxDecayRates2}) we are able to compute the asymptotic state, and prove that the OQS thermalizes to a thermal state with the temperature of RII. Consider the differential equation for the upper population in Eq.~(\ref{eqn:MEmatrix}), and the asymptotic limit
\begin{equation}
	\rho^{\textrm{ss}}=\lim_{t \rightarrow \infty} \rho_\textrm{S}(t)~,
\end{equation}
where the density matrix becomes independent of time, then
\begin{equation}
	\rho_{++}^{\textrm{ss}}=\lim_{t \rightarrow \infty} \frac{\gamma_+(t)}{\gamma_+(t)+\gamma_-(t)}=\frac{e^{-\beta_{\textrm{II}} \omega_0/2}}{e^{\beta_{\textrm{II}} \omega_0/2}+e^{-\beta_{\textrm{II}} \omega_0/2}}~.
\end{equation}
The coherence matrix element obeys an oscillatory decay equation, encoding the decoherence of the OQS. Trace preservation and hermiticity of the density matrix can be used to obtain the remaining matrix elements, so as to check that, indeed
\begin{equation}
	\rho^{\textrm{ss}}=\frac{e^{-H_\textrm{S}\beta_{\textrm{II}}}}{Z(\beta_{\textrm{II}})}~,
\end{equation}
i.e., the OQS asymptotic state is a thermal state at temperature $\beta_{\textrm{II}}$.

\section{Asymtotic limit of the center of the ball of accessible states}\label{append:CenterAymptotic}

To consider the limit $J_{\textrm{II}}(\omega_0) t \to \infty$ of Eq.~(\ref{eqn:CenterSphere}), we first introduce the following change of variables
\begin{equation}
	x = A e^{-J_{\textrm{II}}(\omega_0) t' }~, 
\end{equation}
with
\begin{equation}
	A=2 (n_{\textrm{I}}(\omega_0)-n_{\textrm{II}}(\omega_0))  \frac{J_\textrm{I}(\omega_0)}{J_{\textrm{II}}(\omega_0)}~,
\end{equation}

\begin{equation}
	B = (2 n_{\textrm{II}}(\omega_0)+1)  \frac{J_\textrm{I}(\omega_0)}{J_{\textrm{II}}(\omega_0)}~,
\end{equation}
and
\begin{equation}
	\epsilon = e^{-J_{\textrm{II}}(\omega_0) t}~.
\end{equation}
Then Eq.~(\ref{eqn:CenterSphere}) becomes
\begin{equation}
	c(\epsilon) = \frac{J_\textrm{I}(\omega_0)}{J_{\textrm{II}}(\omega_0)} (A \epsilon)^B e^{A \epsilon}\int_A^{A\epsilon}  x^{-B -1} e^{-x} dx
\end{equation}
which can be rewritten as a difference of two incomplete gamma functions
\begin{equation}\label{eqn:CenterAsymptoticChange}
	c(\epsilon) = \frac{J_\textrm{I}(\omega_0)}{J_{\textrm{II}}(\omega_0)} (A \epsilon)^B e^{A \epsilon} \left[ \Gamma(-B,A)-\Gamma(-B, A\epsilon)\right]~,
\end{equation}
where we made use of the definition
\begin{equation}
	\Gamma(a,z) = \int_z^\infty t^{a-1} e^{-t} dt ~.
\end{equation}
The first term of Eq.~(\ref{eqn:CenterAsymptoticChange}) is null in the limit $\epsilon \to 0$ ($J_{\textrm{II}}(\omega_0) t \to \infty$), while the second one becomes Eq.~(\ref{eqn:CenterAsymptotic}) after taking the limit
\begin{equation}
	\lim_{\epsilon\to 0} \frac{\Gamma(-B, A\epsilon)}{(A\epsilon)^{-B}} = \frac{1}{B}~.
\end{equation}

\section{Non-Equilibrium Asymptotic State}\label{append:NonEqAsymptotic}

The addition of a new environment is encoded, in the ME for the upper population, as
\begin{equation}
	\dot{\rho}_{++}(t)=\sum_{\nu=\{\textrm{L},\textrm{R}\}}\left(\gamma_+^{(\nu)}(t)-\rho_{++}(t)[\gamma_+^{(\nu)}(t)+\gamma_-^{(\nu)}(t)]\right)~,
\end{equation}
which in the steady state limit allows us to obtain
\begin{equation}
	\rho_{++}^{\textrm{ss}}=\lim_{t \rightarrow \infty} \frac{\gamma_+^{\textrm{L}}(t)+\gamma_+^{\textrm{R}}(t)}{\gamma_+^{\textrm{L}}(t)+\gamma_-^{\textrm{R}}(t)+\gamma_+^{R}(t)+\gamma_-^{R}(t)}~.
\end{equation}
By taking the limit, one obtains the following matrix element of the non-equilibrium steady state
\begin{equation}\label{eqn:noneq-asymptotic}
	\rho_{++}^{\textrm{ss}}= \frac{J_I^{\textrm{L}}(\omega_0)n_{\textrm{II}}^{\textrm{L}}(\omega_0)+J_\textrm{I}^{\textrm{R}}(\omega_0)n_{\textrm{II}}^{\textrm{R}}(\omega_0)}{J_\textrm{I}^{\textrm{L}}(\omega_0)(2n_{\textrm{II}}^{\textrm{L}}(\omega_0)+1)+J_\textrm{I}^{\textrm{R}}(\omega_0)(2n_{\textrm{II}}^{\textrm{R}}(\omega_0)+1)}~,
\end{equation}
while the coherence matrix element in the steady state is null. The rest of the matrix elements can be obtained from the trace preserving property of the evolution. Even though this steady state does not directly represent a thermal state, one can obtain an effective temperature for the OQS, since it is a diagonal state in the basis of $H_\textrm{S}$. The effective temperature of the OQS in the steady state is defined as
\begin{equation}
	\beta_{\textrm{eff}}=\frac{1}{\omega_0} \ln \left(\frac{1-\rho_{++}^{\textrm{ss}}}{\rho_{++}^{\textrm{ss}}}\right)~,
\end{equation}
which is a function of both temperatures $\beta_{\textrm{II}}$ of the environments. A very similar expression is obtained for the quasi-stationary states $\rho_{++}^{\textrm{qs}}(\beta_i^\textrm{L},\beta_j^\textrm{R})$ 
\begin{equation}
	\frac{J_\textrm{I}^{\textrm{L}}(\omega_0)n_{i}^{\textrm{L}}(\omega_0)+J_\textrm{I}^{\textrm{R}}(\omega_0)n_{j}^{\textrm{R}}(\omega_0)}{J_\textrm{I}^{\textrm{L}}(\omega_0)(2n_{i}^{\textrm{L}}(\omega_0)+1)+J_\textrm{I}^{\textrm{R}}(\omega_0)(2n_{j}^{\textrm{R}}(\omega_0)+1)}~,
\end{equation}
where the indices $i,j$ refer to whether the corresponding environment is in the prethermal state ($i=\text{I}$) or in the asymptotic state ($i=\text{II}$). This allows us to obtain the quasi-stationary state of the OQS when both environments are prethermalizing ($i=j=\text{I}$), or the state when one has thermalized while the other remains in the prethemalization stage ($i=\text{I}$, $j=\text{II}$, or interchanged). When $i=j=\text{II}$ it corresponds to the asymptotic state of Eq.~(\ref{eqn:noneq-asymptotic}).

\bibliography{biblio}

\end{document}